\documentclass[graybox, nosecnum,sort&compress]{svmult}

\usepackage{amsmath}
\usepackage{mathptmx}       
\usepackage{helvet}         
\usepackage{courier}        
\usepackage{type1cm}        
\usepackage{makeidx}         
\usepackage{graphicx}        
\usepackage{multicol}        
\usepackage[bottom]{footmisc}
\usepackage{hyperref}        
\usepackage{soul}            
\usepackage{color}
\usepackage{subfigure}
\usepackage{booktabs}
\hypersetup{colorlinks=true,urlcolor=blue}
\usepackage[square,numbers]{natbib}
\makeindex             

\begin{document}
\title*{Advances in Machine and Deep Learning for Modeling and Real-time Detection of Multi-Messenger Sources}
\titlerunning{Deep Learning for Multi-Messenger Astrophysics} 
\author{E. A. Huerta \thanks{corresponding author} and Zhizhen Zhao}
\institute{E. A. Huerta \at Data Science and Learning Division, Argonne National Laboratory, Lemont, Illinois 60439,
USA \& Department of Computer Science, University of Chicago, Chicago, Illinois 60637, USA \& Department of Physics, University of Illinois at Urbana-Champaign, Urbana, Illinois 61801, USA. \email{elihu@anl.gov, elihu@uchicago.edu, elihu@illinois.edu}
\and Zhizhen Zhao \at Department of Electrical and Computer Engineering \& Coordinated Science Laboratory \& Department of Mathematics \&  Department of Statistics \& National Center for Supercomputing Applications \& Center for Artificial Intelligence Innovation, University of Illinois at Urbana-Champaign, Urbana, Illinois 61801, USA.  \email{zhizhenz@illinois.edu}}
%
%
\maketitle
\abstract{We live in momentous times. The science community is 
empowered with 
an arsenal of cosmic messengers to study the Universe in unprecedented detail. 
Gravitational waves, electromagnetic waves, neutrinos and cosmic rays cover a wide 
range of wavelengths and time scales. Combining and processing these datasets 
that vary in volume, speed and 
dimensionality 
requires new modes of instrument 
coordination, funding and international collaboration with a specialized human and 
technological infrastructure. In tandem with the advent of large-scale scientific facilities, 
the last decade has experienced an unprecedented transformation in computing 
and signal processing algorithms. The combination of graphics processing units, 
deep learning, and the availability of open source, high-quality datasets, have 
powered the rise of artificial intelligence. This digital revolution now powers 
a multi-billion dollar industry, with far-reaching implications in  
technology and society. In this chapter we describe pioneering efforts to 
adapt artificial intelligence algorithms to address computational grand challenges 
in Multi-Messenger Astrophysics. We review the rapid evolution of these 
disruptive algorithms, from the first class of algorithms 
introduced in early 2017, to the sophisticated algorithms that now 
incorporate domain expertise in their 
architectural design and optimization schemes. We discuss the importance of 
scientific visualization and 
extreme-scale computing in reducing time-to-insight and obtaining new knowledge 
from the interplay between models and data.}
\section{Keywords} 
Multi-Messenger Astrophysics; Numerical Relativity; Gravitational Waves; Electromagnetic Surveys; Big Data; Deep Learning; Artificial Intelligence; Interpretable Artificial Intelligence; High Performance Computing; Real-time Inference
\section{Introduction}

This chapter provides a summary of recent 
developments harnessing the data revolution to 
realize the science goals of Gravitational Wave 
Astrophysics. 
This is an exciting journey that is powered by the 
renaissance of artificial intelligence, and a new 
generation of researchers that are willing to embrace 
disruptive advances in innovative computing and 
signal processing tools. 

In this chapter, machine learning refers to a class of algorithms that 
can learn from data to solve new problems without 
being explicitly re-programmed. While traditional 
machine learning algorithms, e.g., random forests, 
nearest neighbors, etc., have been used successfully in 
many applications, they are limited in their ability to 
process raw data, usually requiring time-consuming 
feature engineering to pre-process data into 
a suitable representation for each application. On 
the other hand, deep learning algorithms can learn 
patterns from unstructured data, finding useful 
representations and 
automatically extracting relevant features for 
each application. The ability of deep learning to deal 
with poorly defined abstractions and problems 
has led to major advances in image recognition, speech, 
computer vision applications, robotics, 
among others~\cite{DL-Book}.

The following sections describe a few 
noteworthy applications of 
modern machine learning for gravitational wave modeling, 
detection and inference. It is the expectation that by 
the time this chapter is published, 
the ongoing developments at the interface of artificial 
intelligence and extreme-scale computing will have 
leapt forward, making this chapter a reminiscence of a 
fast-paced, evolving field of research. The chapter 
concludes with 
a summary of recent applications at the interface of 
deep learning and 
high performance computing to address computational grand 
challenges 
in Gravitational Wave Astrophysics.

\section{\textit{Machine learning and numerical relativity for gravitational wave source modeling}}
\label{sec:model}

One of the first examples of gravitational wave source modeling was 
introduced by Einstein. He derived an approximated version of 
his field equations~\cite{gr} to confirm that general relativity 
accurately predicts the precession of the perihelion of Mercury~\cite{kenbook}. 
Shortly after Einstein published his general theory of relativity, 
Karl Schwarzschild found an exact solution to Einstein's field equations, 
known as the Schwarzschild metric~\cite{Schwarzschild:1916uq}. This analytical solution 
describes the gravitational field outside of a spherical mass that has no charge and no spin, under the assumption that the cosmological constant is zero. Soon afterwards, Reissner and Nordstr\"{o}m 
derived an analytical 
solution that describes the gravitational field exterior to a charged, 
non-spinning spherical mass~\cite{1916Reissner}. Nearly five decades later, and 
with the understanding that these metrics describe the gravitational field 
outside black holes, Roy Kerr discovered the analytical solution that 
describes uncharged, spinning black holes~\cite{kerr}. Shortly thereafter, the Kerr metric was extended to the case of charged, spinning black holes---the 
Kerr-Newman metric~\cite{1965KerrNewman}.

While these analytical solutions provided tools to 
extract new insights from general relativity, 
there were astrophysical scenarios of interest that required 
novel approaches. The development of approximate solutions to 
Einstein's equations, such as the post-Newtonian~\cite{Blanchet:2013haa} and post-Minkowskian 
formalisms~\cite{pmdamour}, provided a better understanding 
of gravitationally bound systems like neutron star mergers that 
were prime targets for gravitational wave detection. Still, a detailed study of 
gravitational wave emission in the strong, highly dynamical gravitational field of
black hole mergers required a complete numerical solution of 
Einstein's field equations. In the late 1990s, and 
after decades of mathematical 
and numerical developments, the Binary Black Hole Grand 
Challenge Alliance, funded by the US National Science Foundation, successfully simulated a head-on binary black hole collision~\cite{1995Sci...270..941M}. 
The first successful evolution of binary black hole spacetimes, 
including calculations about the orbit, merger, and gravitational waves emitted, were reported in~\cite{preto}. Afterwards, other numerical 
relativity teams reported similar accomplishments~\cite{camp:2006,baker:2006}. 

Within a decade, numerical relativity 
matured to the point of harnessing high performance computing 
with mature software stacks to study the physics and gravitational
wave emission of binary black hole mergers covering a wide range 
of astrophysical scenarios of interest~\cite{Mroue:2013xna,Jani:2016wkt,Healy:2020vre,huerta_nr_catalog}. These resources have been 
used extensively to develop semi-analytical waveform models that 
describe the inspiral-merger-ringdown of binary black hole 
mergers~\cite{Hannam:2013oca,Bohe:2016gbl,Khan:2015jqa} and to inform the design of algorithms for gravitational wave detection~\cite{Aylott:2009ya}. 

\begin{figure*}
\begin{center}
    \includegraphics[width=0.9\textwidth]{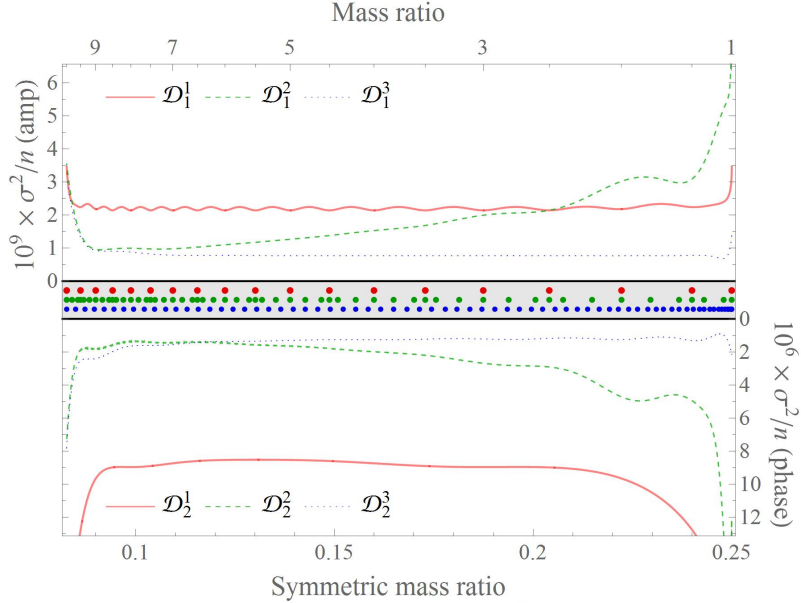}
  \caption{Interpolation error, $\sigma$, of a Gaussian Process Emulator on the amplitude (top panel), and phase (bottom panel), for three training datasets that describe quasi-circular, non-spinning binary black hole mergers: $\mathcal{D}^{1}_{\alpha}$ (red), $\mathcal{D}^{2}_{\alpha}$ (green), and $\mathcal{D}^{3}_{\alpha}$ (blue). Each numerical relativity waveform in these sets has $n\!=\!2800$ time samples. The (symmetric) mass ratio values for the points in $\mathcal{Q}^{1}$, $\mathcal{Q}^{2}$ and $\mathcal{Q}^{3}$ are indicated respectively by red, green, and blue dots along the central horizontal panel. Each iteration of the training set reduces the error in the amplitude and phase interpolations. \texttt{This Figure was produced by the authors of this chapter in the published article~\cite{ENIGMA_Huerta}.}\label{fig:gpe}}
\end{center}
\end{figure*}

In time, the need to produce accurate and computationally efficient waveform models became apparent. This need has led to the adoption of surrogate models~\cite{PhysRevD.95.104023,Rifat:2019ltp,Lackey:2018zvw,Varma:2018mmi} and traditional machine learning techniques, such as Gaussian emulation~\cite{ENIGMA_Huerta,PhysRevD.101.063011}. 
The development of fast waveform generators has had a significant impact in gravitational wave parameter estimation studies. This is because Bayesian 
parameter estimation that utilizes Markov-chain Monte Carlo requires \({\cal{O}}\left(10^8\right)\) waveform evaluations. Surrogate 
waveforms are tailored for this task, since each 
waveform may be produced within 50ms with typical mismatches of order \(10^{-3}\) with other state-of-the-art waveform approximants~\cite{PhysRevD.95.104023}. Fig.~\ref{fig:gpe} shows an example of a Gaussian emulator used to create a stand-alone merger waveform trained with a catalog of quasi-circular, non-spinning, binary black hole mergers. Notice how the performance of the emulator changes depending on the format with which the training dataset is presented to the emulator. This is a key difference between traditional machine learning and deep learning, since in the latter case, neural networks do not need feature engineering to provide optimal performance.

Large-scale numerical relativity waveform catalogs have 
also been used as 
datasets to train Gaussian emulators to produce highly accurate modeled waveforms in record time~\cite{ENIGMA_Huerta}. It is also known that traditional machine learning methods, such as surrogate models or Gaussian emulation, 
present several challenges when trying to organize data in high-dimensional spaces~\cite{2017PhRvD..96b4058B}. In stark contrast, deep neural networks excel at learning high-dimensional data and provide additional advantages such as improving convergence and performance when combined with extreme-scale computing. 

While rapid progress has taken place in the modeling of binary black hole mergers, matter systems such as binary neutron star mergers and 
black hole-neutron star mergers continue to present significant challenges~\cite{Foucart:2018lhe}. 
It is expected that numerical relativity will meet these challenges within 
the next few years with the production of open source numerical relativity software that will bring together experts across the community. The adoption of deep learning to accelerate the description of physics that requires sub-grid scale precision, such as turbulence, is also in earnest development~\cite{2020PhRvD.101h4024R}.  

In summary, numerical relativity has played a key 
role in the development of waveform models and signal processing 
algorithms that enabled 
the discovery of gravitational waves. Numerical relativity and machine 
learning have been combined to produce modeled waveforms 
for production-scale gravitational wave analyses. We also see a new 
trend in which numerical relativity and deep learning have been combined for 
waveform production at scale. This approach may well provide a solution for 
high-dimensional signal manifolds. Another exciting trend is the use 
of deep learning to replace compute-intensive modules in numerical 
relativity software that describes matter systems.

\section{\textit{Machine learning for gravitational wave data analysis}}
\label{sec:gwda}

In this section we review machine learning applications in the context of 
parameter estimation, rapid characterization of compact binary remnants, 
and signal denoising.

\noindent \textbf{Parameter Estimation} Machine learning 
applications for 
gravitational wave inference have been developed to overcome 
the computational expense and 
poor scalability of established Bayesian approaches. 
Estimating the posterior probability density functions 
of astrophysical parameters that describe gravitational wave sources 
is a computationally intensive task. This is because these systems 
span a 15-D parameter space, thereby requiring a large number of modeled waveforms 
to densely sample this signal 
manifold. In the previous section we discussed the importance of harnessing machine learning methods to produce modeled waveforms at scale. 
In addition to massive waveform generation, low to moderate signal-to-noise ratios and complex noise anomalies 
may require additional follow-up studies, thereby demanding additional
computational resources. In order to mitigate these computational challenges,
parameter estimation algorithms such as \texttt{LALInference}~\cite{bambiann:2015PhRvD} and \texttt{PyCBC Inference}~\cite{biwer:2018osg} use nested 
sampling~\cite{article_nested} or Markov Chain Monte Carlo~\cite{mcmc_paper}. These techniques usually 
take days to weeks to produce posterior samples of gravitational wave sources' 
parameters. Thus, it is timely and relevant to explore new approaches to 
further reduce time-to-insight. This is ever more urgent as the international 
network of gravitational wave detectors continues to increase its 
sensitivity, thereby increasing the detection rate of observed events. 

Machine learning solutions to accelerate parameter estimation include 
Gaussian process emulation~\cite{Lange:2018pyp}, nested sampling~\cite{Feroz:2009de}, and 
nested sampling combined with neural networks~\cite{bambi:2012MNRAS}. In the latter case, 
likelihood calculations are accelerated by up to 100x for computationally 
demanding analyses that involve long signals, such as neutron star systems.

\noindent \textbf{Rapid characterization of compact binary remnants} 
Multi-Messenger searches demand real-time detection of 
gravitational wave sources, accurate sky localization, and information 
regarding the nature of the source, in particular whether the progenitor 
may be accompanied by electromagnetic or neutrino counterparts. 
To address the latter point, i.e., to ascertain whether the 
remnant is a neutron star, or whether in the case of 
a neutron star-black hole merger the black hole remnant is surrounded by an 
accretion disk of tidally disrupted material from the neutron star, it may be possible to use information 
provided by low-latency detection algorithms. However, it is known that 
these estimates may differ from accurate but  hours- to days-long parameter estimation studies. 
To start addressing these limitations, a supervised nearest neighbors 
classification method was introduced 
in~\cite{Chatterjee:2019avs}. This method infers, in a fraction of a second, 
whether a compact binary merger will have an 
electromagnetic counterpart, thus providing time-critical information 
to trigger statistically informed electromagnetic follow-up searches.

On the other hand, the ever-increasing catalog of detected 
gravitational wave sources provides the means to infer the 
mass and spin distributions of stellar mass compact binary systems. 
These studies will shed new light on the stellar evolution processes 
that may lead to the formation of these astrophysical objects, or whether 
these objects are formed from a mixture of different populations~\cite{Mandel:2016prl}. 
Agnostic studies that involve Gaussian mixture models are ideally suited 
to enable data-driven analyses~\cite{Powell:2019nmw}.

\noindent \textbf{Signal denoising} Gravitational wave signals are 
contaminated by environmental and instrumental noise sources that 
are complex to model and difficult to remove. Several methods 
have been explored for data cleaning and 
noise subtraction, ranging from the basic Wiener filter that optimally 
removes linear noise~\cite{wiener_1643650} to machine learning applications that can 
effectively remove linear, non-linear, non-Gaussian and non-stationary 
noise contamination~\cite{PhysRevD.101.042003,Cavaglia_2019}. Machine learning methods 
employed to remove noise from gravitational wave signals includes 
Bayesian methods~\cite{Cornish:2014kda}, 
dictionary learning~\cite{Torres-Forne:2016dwq,PhysRevD.102.023011}, principal component 
analysis~\cite{Heng_2009}, and total-variational methods based on \(L_1\) 
norm minimization techniques that were originally developed in the context of 
image processing, but were subsequently adapted to clean signals in time- and 
frequency-domain~\cite{total_var_font}.

This succinct summary of traditional machine learning applications in 
gravitational wave astrophysics suggests that part of this community 
has been engaged in harnessing advances in computing and 
signal processing to address computational grand challenges in 
this field. The following section shows how this process has been 
accelerated with the rise of artificial intelligence from the early 
2010s.

\section{\textit{Deep Learning for gravitational wave data analysis}}

Gravitational wave data analysis encompasses a number of core tasks, 
including detection, parameter estimation, data cleaning, glitch 
classification and removal, and signal denoising. In this section we 
present a brief overview of the rapid rise of artificial intelligence for 
gravitational wave astrophysics.

\noindent \textbf{Detection} Existing algorithms for signal detection 
include template matching, where the physics of the source is 
used to inform the classification of noise triggers, and to 
identify those that describe gravitational wave 
sources~\cite{2016CQGra..33u5004U,Sachdev:2019vvd}. Other 
events where the underlying astrophysics is unknown or too 
complex to capture in modeled waveforms take advantage of burst searches, 
which make minimal assumptions about the morphology of 
gravitational wave signals~\cite{PhysRevD.93.042004}. 
Continuous wave sources, 
such as isolated 
neutron stars, emit signals that, although well known, are 
very weak and long. This combination makes their search and detection 
very computationally intensive~\cite{Riles:2017evm}.

As mentioned above, as advanced LIGO and the international gravitational
wave detector 
network gradually reach design sensitivity, core data analysis studies 
will outstrip the capabilities 
of existing computing facilities if we continue to use poorly scalable 
and compute-intensive signal processing methods. Furthermore, gravitational 
wave astrophysics is not the only discipline with an ever-increasing 
need for computing resources. The advent of other large-scale scientific 
facilities such as the Square Kilometer Array, the High Luminosity 
Large Hadron Collider, or the Legacy Survey of Space and Time, to mention 
a few ~\cite{ska_ieee,ApollinariG.:2017ojx,lsst}, will produce datasets 
with ever-increasing complexity 
and volume. Thus, a radical approach in terms of computing and signal processing is 
needed to maximize and accelerate scientific discovery in the big data era.

To contend with these challenges, a disruptive approach that 
combines deep learning and high performance computing was introduced 
in~\cite{geodf:2017a}. This idea was developed to address a number of specific 
challenges. To begin with, the size of modeled waveform catalogs used for 
template matching searches imposes restrictions on the science reach 
of low-latency searches. Thus, it is worth exploring a different 
methodology that enables real-time gravitational wave detection without 
sacrificing the depth of the signal manifold that describes astrophysical 
sources. It turns out that there is indeed a signal processing tool, deep 
learning~\cite{DeepLearning}, that encapsulates information in 
a hierarchical manner, 
bypassing the need to use large catalogs of images or time-series for accelerated inference. The second key consideration in the use of deep learning 
for gravitational wave detection is the fact that real signals may be 
located anywhere in the data stream broadcast by detectors. 
Thus, the neural 
networks in~\cite{geodf:2017a,GEORGE201864} introduced 
the concept of time-invariance. A third 
consideration is that there is no way to predict the 
signal-to-noise ratio of 
real events. Thus, the methods presented in~\cite{geodf:2017a,GEORGE201864} showed how 
to adapt curriculum learning~\cite{curriculum_learning}, originally developed in the context of 
image processing, to do classification or detection of noisy and weak 
signals. The key idea behind this approach consists of 
training the model by first exposing it to signals with high 
signal-to-noise ratio, and then gradually increasing the noise content until 
the signals become noise-dominated. The combination of the aforementioned innovations led to the 
realization that neural networks could indeed detect modeled 
gravitational wave signals embedded in simulated advanced LIGO noise 
with the same sensitivity as template matching algorithms, but orders 
of magnitude faster with a single, inexpensive GPU. In addition to these 
results, the authors in~\cite{geodf:2017a} showed how to modify a neural network 
classifier and use transfer learning to construct a neural network 
predictor that provides real-time point-parameter estimation results for 
the masses of the binary components, mirroring 
a similar capability of established low-latency detection analyses. 
This seminal work was then extended to the case of real 
gravitational wave signals in advanced LIGO noise~\cite{GEORGE201864}. About a year 
later, different teams reported similar classification results 
in the context of modeled signals in simulated LIGO noise~\cite{2018GN}. 

A metric for the impact of the seminal ideas laid out 
in~\cite{geodf:2017a,GEORGE201864} is given by the number of research teams across the 
world that have reproduced and extended these 
studies~\cite{2018GN,2020arXiv200914611S,Lin:2020aps,Wang:2019zaj,Nakano:2018vay,Fan:2018vgw,Li:2017chi,Deighan:2020gtp}. It is also 
worth mentioning, however, that while these studies demonstrated the 
scalability, computational efficiency and sensitivity of neural 
networks for gravitational wave detection, it is still essential 
to demonstrate the use of these signal processing tools for searches that 
span a high-dimensional signal manifold, and to apply them 
to process large datasets.

Deep learning has been applied to the detection of neutron star mergers~\cite{Miller:2019jtp,Krastev:2019koe,2020PhRvD.102f3015S}, forecasting of neutron stars inspirals and neutron star-black hole mergers~\cite{Wei:2020sfz,2020arXiv201203963W}, continuous wave 
sources~\cite{Dreissigacker:2020xfr,Dreissigacker:2019edy,2020PhRvD.101f4009B}, signals with complex morphology~\cite{2020arXiv200914611S}, and to 
accelerate waveform production~\cite{Khan:2020fso,PhysRevLett.122.211101}. 
The rapid progress and maturity that these algorithms have achieved 
within just three years, at the time of writing this book, 
suggest that production-scale 
deep learning methods are on an accelerated track to 
become an integral part of gravitational wave discovery~\cite{2021PhLB..81236029W,2020arXiv201208545H}.

\begin{figure}
		\centering     
		\includegraphics[width = \textwidth]{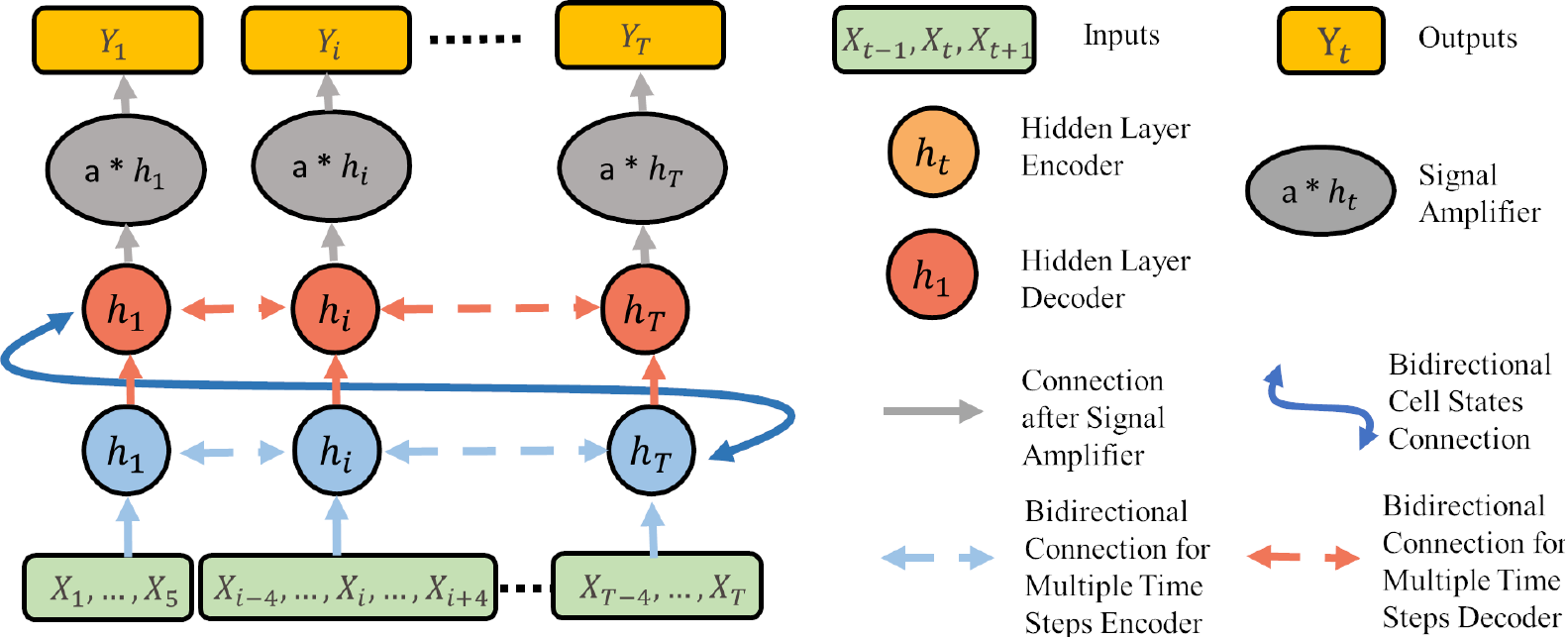}	
		\caption{Architecture of the Enhanced Deep Recurrent Denoising Auto-Encoder model. Double arrows indicate the bidirectional long short-term memory (LSTM) connections. The one-directional arrows represent conventional LSTM structure and the solid blue double arrow connects cross layer cell states at the initial and the last steps in the time window. The last layer is the signal amplifier indicated by a gray ellipse. \texttt{This Figure was produced by the authors of this chapter in the published article~\cite{shen2019denoising}.}}
		\label{fig:structure}
\end{figure}

\noindent \textbf{Signal Denoising} The first deep learning 
application for the removal of noise and noise anomalies 
for gravitational wave signal processing was introduced in~\cite{shen2019denoising}. 
This study described how to combine recurrent neural networks 
with denoising auto-encoders to clean up modeled waveforms 
embedded in real advanced LIGO noise. The different 
components of this Enhanced Deep Recurrent Denoising 
Auto-Encoder (EDRDAE) are
shown in Figure~\ref{fig:structure}. To provide optimal 
denoising performance for low signal-to-noise ratio signals, 
this model incorporated a 
signal amplifier layer, and was trained with curriculum learning. 
Another 
feature of this model is that while it was originally trained to 
denoise signals that describe quasi-circular, non-spinning 
black hole mergers, this model was able to generalize to 
signals that describe eccentric, non-spinning black hole 
mergers, whose morphology is much more complex than the training 
dataset. This study showed that deep learning approaches 
outperform traditional machine learning methods such as 
principal component analysis and dictionary learning for 
gravitational wave signal denoising~\cite{shen2019denoising}.

\begin{figure}
\centerline{
\includegraphics[width=0.8\textwidth]{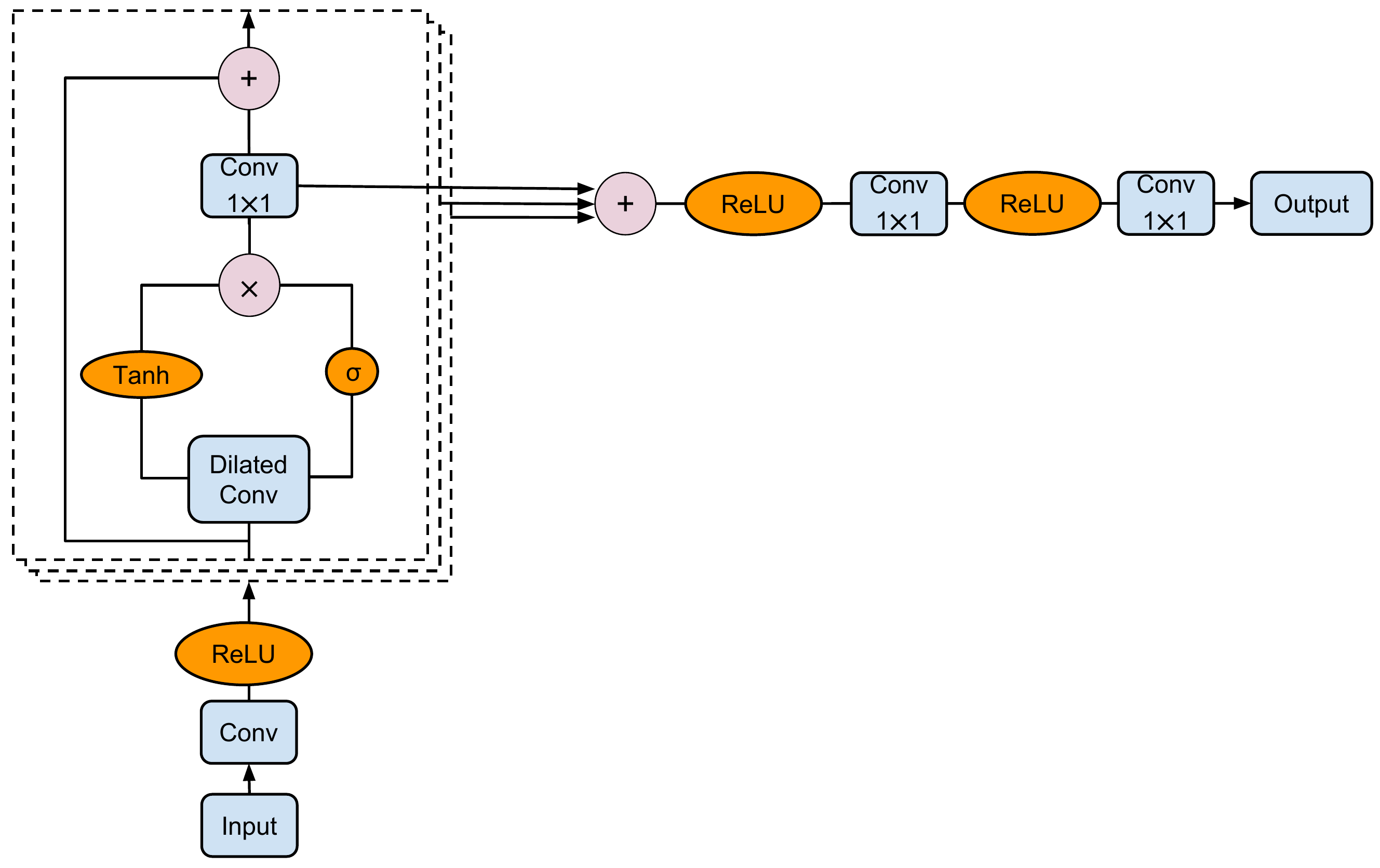}\hspace{-5.5cm}
\raisebox{-0.3cm}{
\includegraphics[width=0.5\textwidth]{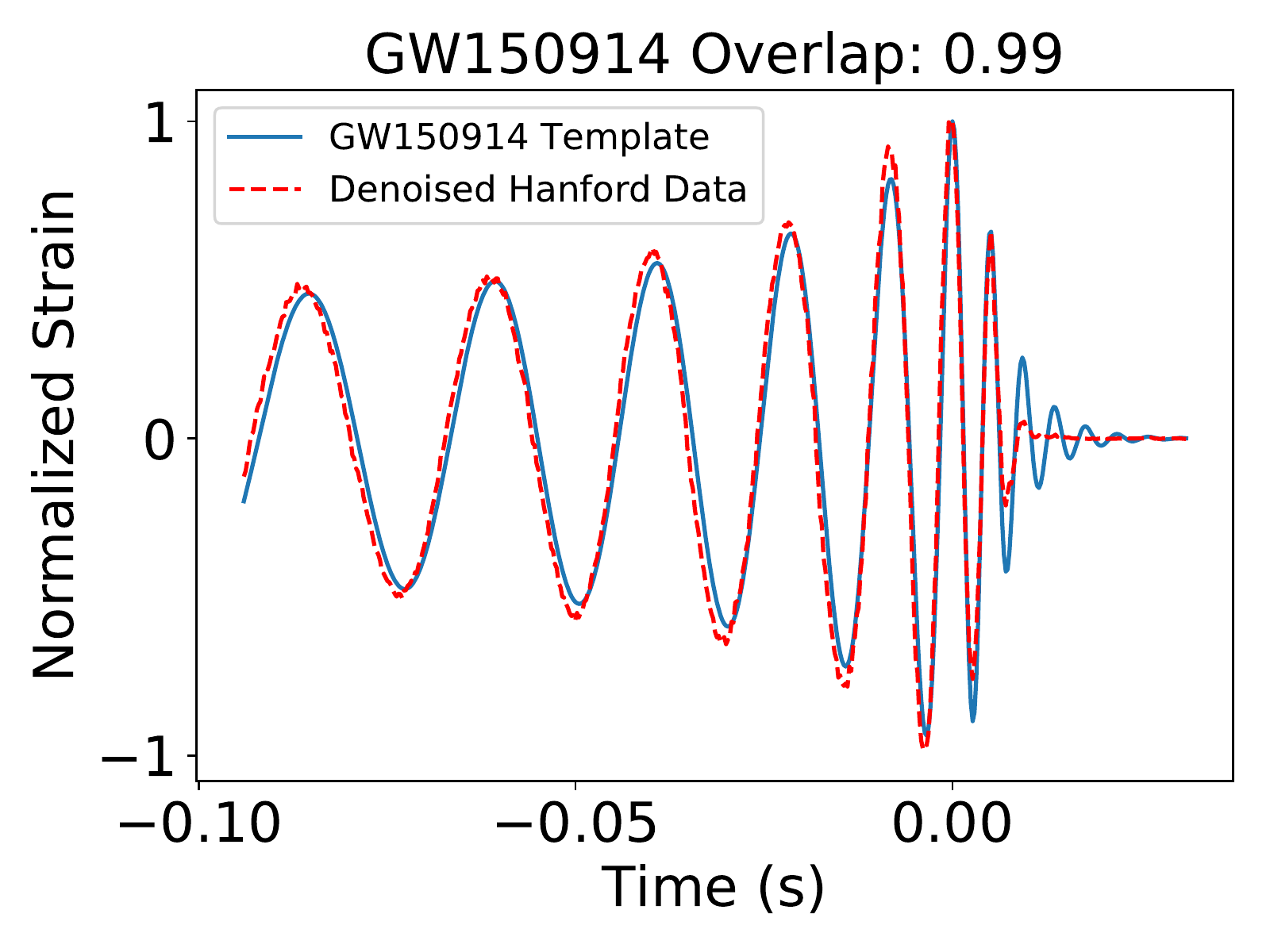}
}
} 
\caption{Left panel: architecture for the denoising \texttt{WaveNet} algorithm. Bottom-right panel: overlap between the output of the \texttt{WaveNet} denoising algorithm upon processing real advanced LIGO data that contains the event GW150914, and the template that best describes this astrophysical source according to matched-filtering algorithms. \texttt{This Figure was produced by the authors of this chapter in the published article~\cite{Wei:2019zlc}.}}
\label{fig:net}
\end{figure}

\noindent The first application of deep learning for 
signal denoising and de-glitching of actual gravitational 
wave observations was introduced in~\cite{Wei:2019zlc}. The model 
proposed for this analyses consists of a repurposed 
\texttt{WaveNet} architecture---see left panel of Figure~\ref{fig:net}---which was originally developed 
for forecasting and human speech generation~\cite{2016wavenet}. The data 
used to train this network consists of one-second-long 
time-series modeled waveforms, sampled at 8192Hz, that describe quasi-circular, 
non-spinning binary black hole mergers. Upon encoding 
time- and scale-invariance, this model was used to 
denoise several gravitational wave signals, as shown in the right panel of Figure~\ref{fig:net},
and to demonstrate 
its effectiveness at removing noise and glitches from simulated signals embedded 
in real advanced LIGO noise. This model was also used to 
denoise quasi-circular, spinning, precessing binary black 
hole mergers, furnishing evidence for the ability of the 
model to generalize to new types of signals that were not used 
in the training stage.

\noindent \textbf{Data cleaning} Recent developments for 
data cleaning include~\cite{PhysRevResearch.2.033066}, 
in which deep learning is applied to gravitational 
wave detector data and data from on-site sensors 
monitoring the instrument to reduce the noise in the 
time-series due to instrumental artifacts and environmental contamination. This approach is able to remove linear, 
nonlinear, and non-stationary coupling mechanisms, improving 
the signal-to-noise ratio of injected signals by up 
to \(\sim 20\%\).

\noindent \textbf{Parameter Estimation} Uncertainty 
quantification 
is a rapidly evolving area in deep learning research. 
Thus, it is natural that a number of methodologies 
have been investigated to constrain the astrophysical 
parameters of gravitational wave sources. For instance, 
in ~\cite{Shen:2019vep}, Bayesian  neural networks were used to 
constrain the astrophysical properties of real 
gravitational wave sources before and after the merger 
event, showcasing the ability of neural networks to 
measure the final mass and spin remnant sources by directly 
processing real LIGO data. Conditional variational 
auto-encoders~\cite{Gabbard:2019rde} and multivariate Gaussian posterior models~\cite{Chua:2019wwt} have been used to construct posterior distributions of modeled 
signals embedded in simulated LIGO noise. In~\cite{Green:2020hst}, the 
authors introduce the use of auto-regressive normalizing 
flows for rapid likelihood-free inference of binary black 
hole mergers that describe an 8-D parameter space. This analysis, 
originally applied for modeled signals in stationary Gaussian noise, 
was extended to cover the 15-D parameter space for GW150914~\cite{Green:2020dnx}. Deep learning has also been explored 
to characterize compact binary populations~\cite{Wong:2020wvd}.

\noindent \textbf{\textit{Deep Learning for the detection and characterization of higher-order waveform multipole signals of eccentric binary black hole mergers}}
\label{sec:HOwaveform}

It has been argued in the literature that gravitational wave 
observations of eccentric binary black hole mergers will provide 
the cleanest evidence of the existence of compact binary populations 
in dense stellar environments, such as galactic nuclei and core-collapsed globular clusters~\cite{PhysRevD.97.103014}.

\begin{figure}
\centerline{
\includegraphics[width=.34\textwidth]{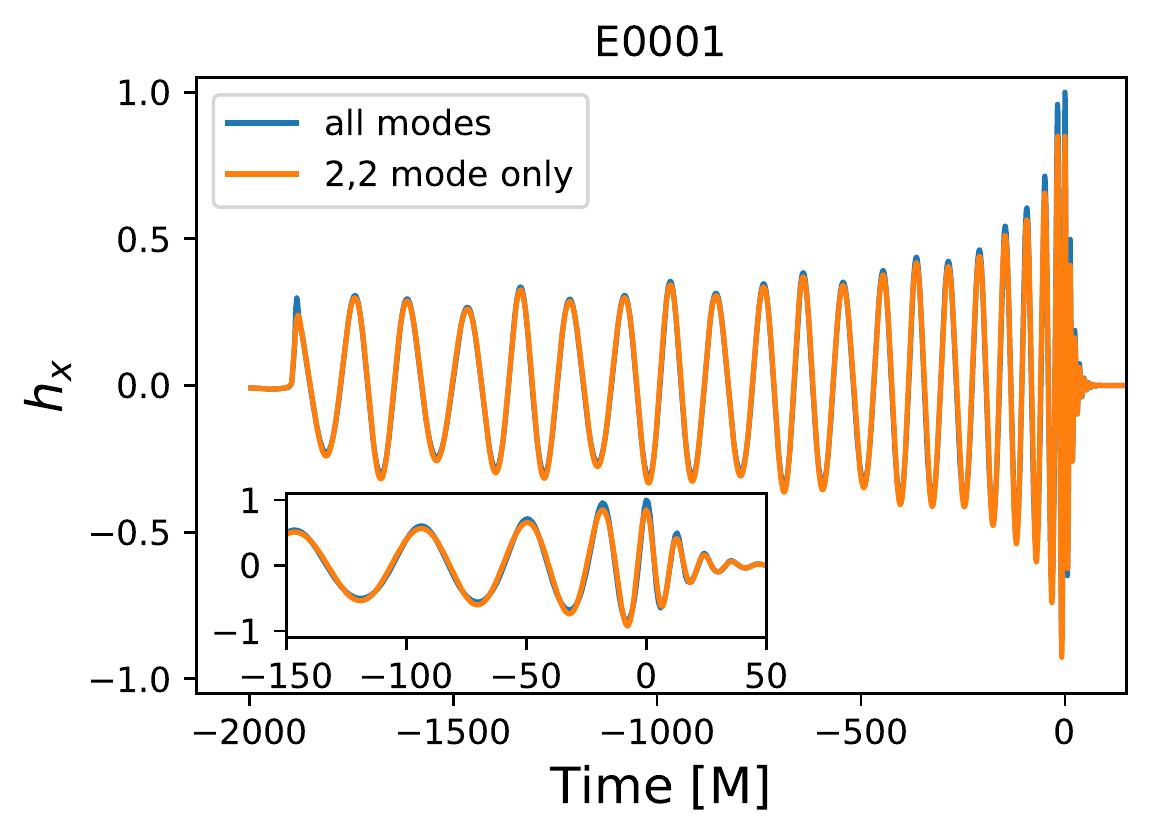}\hspace{.1em}%
\includegraphics[width=.34\textwidth]{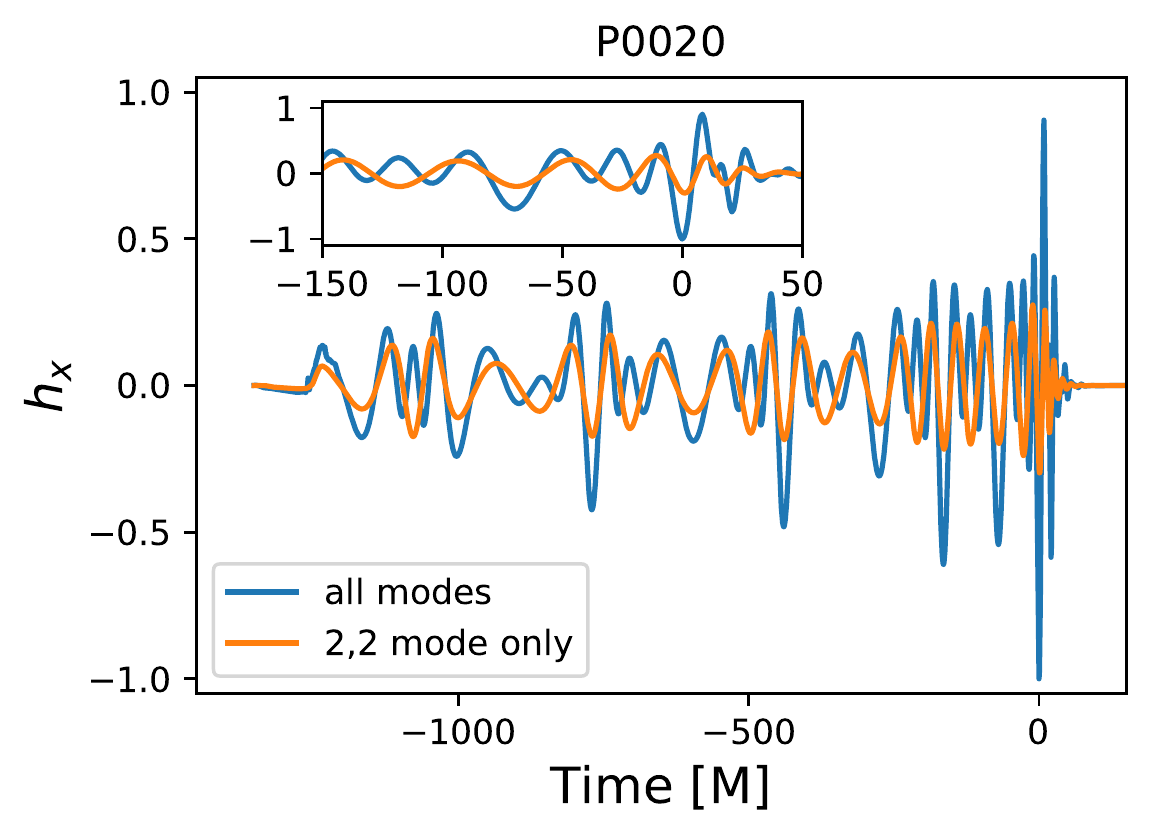}\hspace{0.1em}%
\includegraphics[width=.34\textwidth]{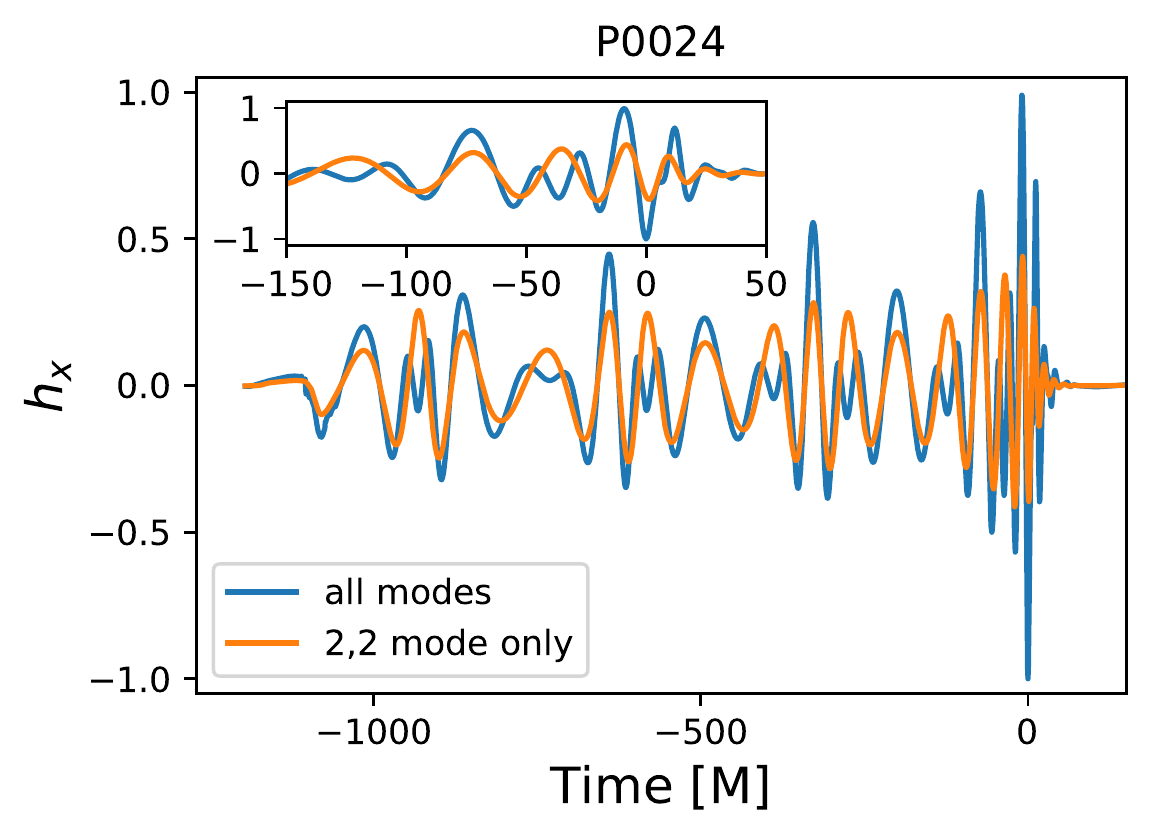}\hspace{0.1em}%
}
\centerline{
\includegraphics[width=.34\textwidth]{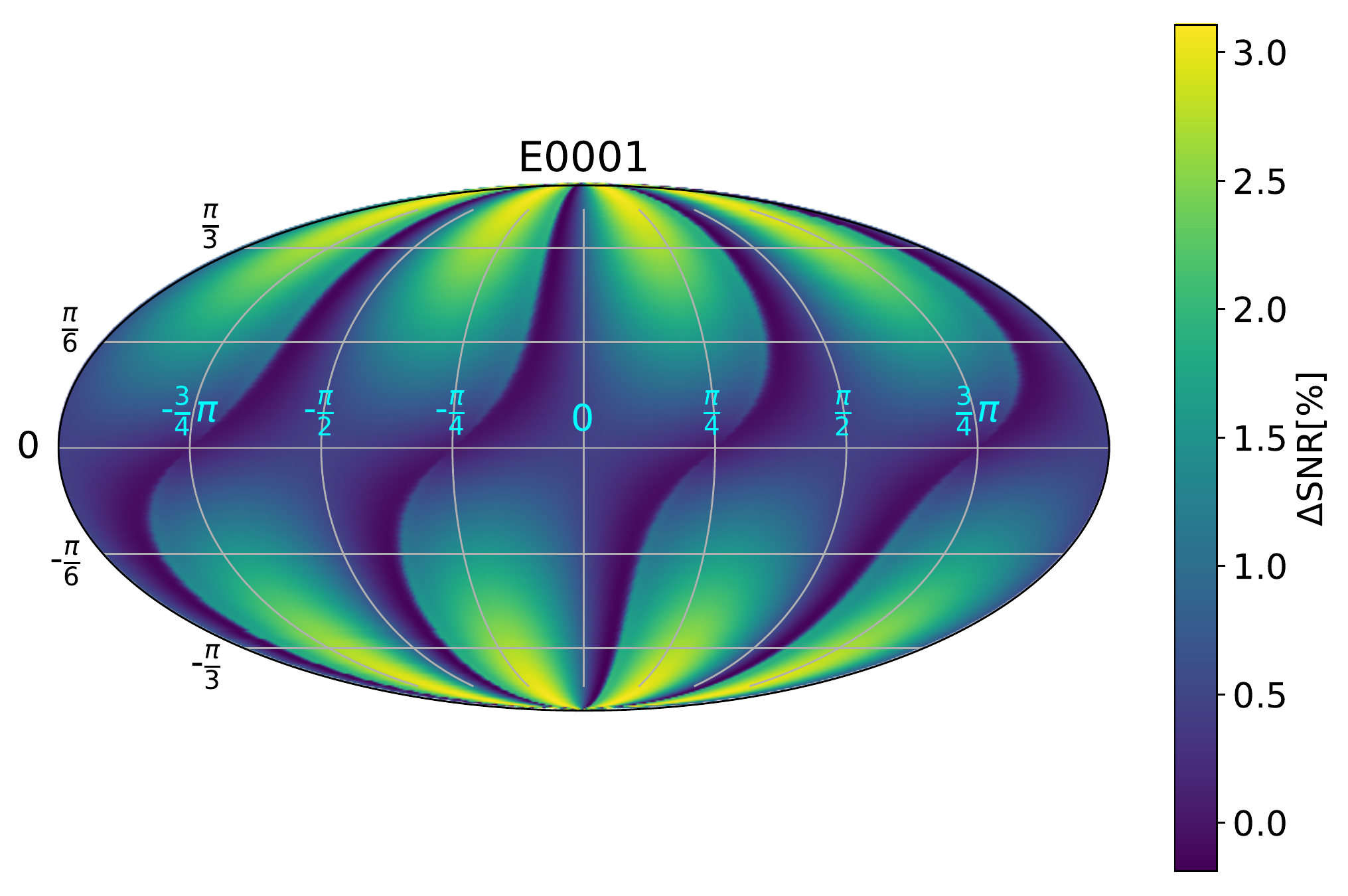}\hspace{0.1em}%
\includegraphics[width=.34\textwidth]{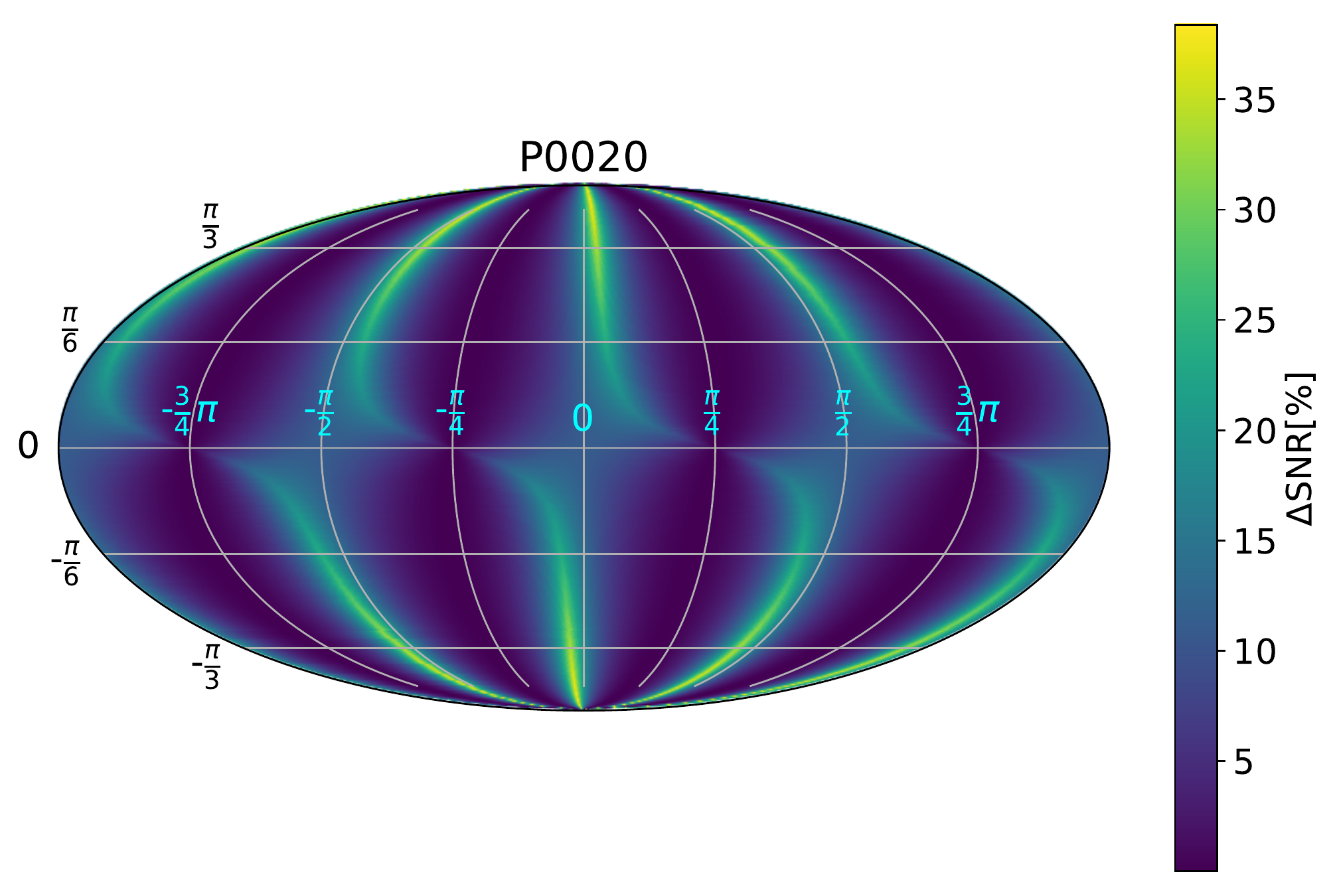}\hspace{0.1em}%
\includegraphics[width=.34\textwidth]{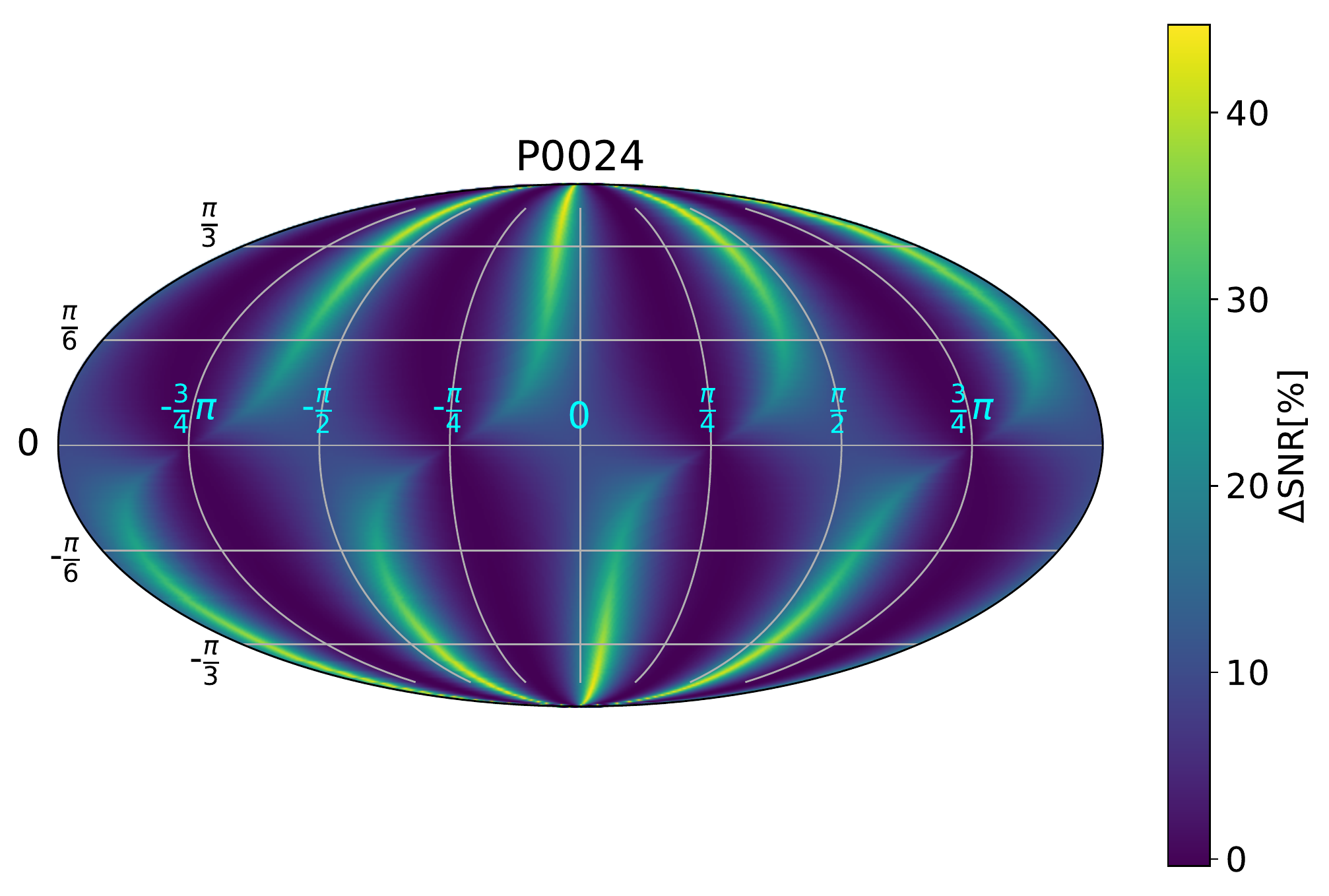}\hspace{0.1em}%
}
\caption{Top panels: comparison between numerical relativity waveforms that include either all \((\ell, |m|)\) modes or the  \(\ell=|m|=2\) mode only, using \((\theta,\,\phi)\) values that maximize the inclusion of higher-order modes in terms of signal-to-noise ratio calculations. Bottom panels: increase in signal-to-noise ratio, \(\Delta \mathrm{SNR}\), due to the inclusion of higher-order modes. These sky distributions are produced with the waveforms shown in the top panels. \texttt{This Figure was produced by the authors of this chapter in the published article~\cite{Adam:2018prd}.}}
\label{fig:snrs_hom}
\end{figure}

The importance of including higher-order waveform modes for 
the detection of eccentric binary black hole mergers has been 
studied in the literature~\cite{Adam:2018prd}. It has been found that, as 
in the case of quasi-circular mergers, higher-order modes play a 
significant role in the detection of asymmetric binary black 
hole mergers~\cite{Prayush:2013a}. For instance, Figure~\ref{fig:snrs_hom} 
shows the 
increase in signal-to-noise ratio due to the inclusion of 
higher-order modes for a variety of astrophysical scenarios. 
These results show that for comparable mass ratio systems, 
represented by the numerical relativity waveform \texttt{E0001} (see Table~\ref{tab:sims}), higher-order modes do not alter the amplitude of the \(\ell=|m|=2\) mode, thereby having a negligible contribution on the signal-to-noise ratio of these systems. However, 
for the asymmetric mass ratio systems represented by 
\texttt{P0020} and \texttt{P0024}, the inclusion of higher-order 
modes leads to a significant increase in the signal-to-noise ratio of these systems.

 \begin{table}
\caption{\label{results} \((q,\,e_0,\, \ell_0,\, x_0)\) represent the mass ratio and the measured values of eccentricity, mean anomaly, and dimensionless orbital frequency parameters.}
		\footnotesize
		\begin{center}
                        \setlength{\tabcolsep}{12pt} 
			\begin{tabular}{c c c c c c}
				\hline 
				Simulation&$q$ & $e_0$ & $\ell_0$ & $x_0$ \\ 
				\hline
				E0001	&	1	&	0.052	&	3.0	&	0.0770	\\
				J0040	&	1	&	0.160	&	3.0	&	0.0761	\\
				L0016	&	5	&	0.140	&	2.9	&	0.0862	\\
				P0016	&	6	&	0.160	&	2.8	&	0.0900	\\
				P0020	&	8	&	0.180	&	2.9	&	0.0936	\\
				P0024	&	10	&	0.180	&	3.0	&	0.0957	\\
				\hline 
			\end{tabular}
		\end{center}
	\label{tab:sims}
	\end{table}
	\normalsize

In Figure~\ref{fig:snrs_hom} the signal-to-noise ratio 
distributions are presented as a function of the source's sky 
location, \((\alpha,\,\beta\)), mapped into a Mollweide 
projection: \((\vartheta, \varphi) \rightarrow (\pi/2-\alpha, \beta-\pi)\). The reference frame \((\theta\,,\phi)\) is anchored 
at the center of mass of the binary system, and determines 
the location of the detector. In this 
reference frame, \(\theta=0\) coincides with the total angular 
momentum of the binary, and \(\phi\) indicates the azimuthal 
direction to the observer. Furthermore, the top panels in 
Figure~\ref{fig:snrs_hom} show that the inclusion of \((\ell, \, |m|)\) modes significantly modifies the ringdown evolution of \(\ell=|m|=2\) waveforms. The finding is in line with studies that indicate the need to include \((\ell, \, |m|)\) modes for tests of general relativity using ringdown waveforms~\cite{BertiSes:2016PRL,YagiSCQG:2016}. 

\begin{figure*}
  \includegraphics[width=0.5\textwidth]{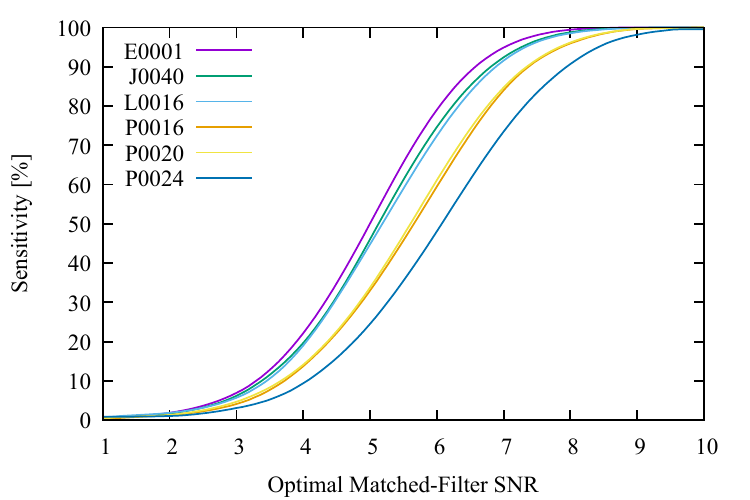}
  \includegraphics[width=0.5\textwidth]{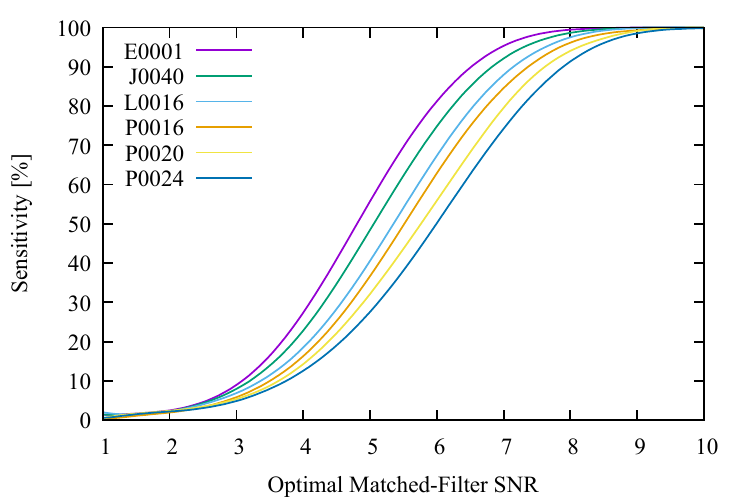}
   \caption{Left panel: neural networks may detect eccentric signals that include higher-order modes with optimal sensitivity in simulated Gaussian noise when their matched-filtering signal-to-noise ratio is \(\geq 10\). Right panel: same as the left panel, but now using real advanced LIGO noise. \texttt{This Figure was produced by the authors of this chapter in the published article~\cite{Adam:2018prd}.}}
 \label{fig:sensitivity_hom}
 \end{figure*}

Having identified a collection of waveforms in which the 
inclusion of higher-order modes induce the most significant 
modification to the \(\ell=|m|=2\) mode, corresponding to the 
maximum gain in signal-to-noise ratio, the authors 
in~\cite{Adam:2018prd} injected these signals into simulated and 
real advanced LIGO noise, and used neural networks to search for 
them. Their findings are shown in Figure~\ref{fig:sensitivity_hom}. 
In a nutshell, deep learning models can identify these complex 
signals, even though they were trained with 
quasi-circular waveforms. Future studies may explore 
whether neural networks improve their sensitivity when 
they are trained with datasets that describe eccentric mergers.

\noindent \textbf{\textit{Deep Learning for the characterization 
of spin-aligned binary black hole mergers}}

Deep learning has been used to study the properties of the 
gravitational wave signal manifold that describes 
quasi-circular, spinning, 
non-precessing binary black hole mergers~\cite{KHAN2020135628}. 
This study explored how neural networks handle parameter 
space degeneracies, and their ability to measure the individual 
spins, effective spin and mass ratio of black hole mergers by 
directly processing waveform signals in the absence of noise. 

\begin{figure}
\centerline{
\includegraphics[width=0.9\linewidth]{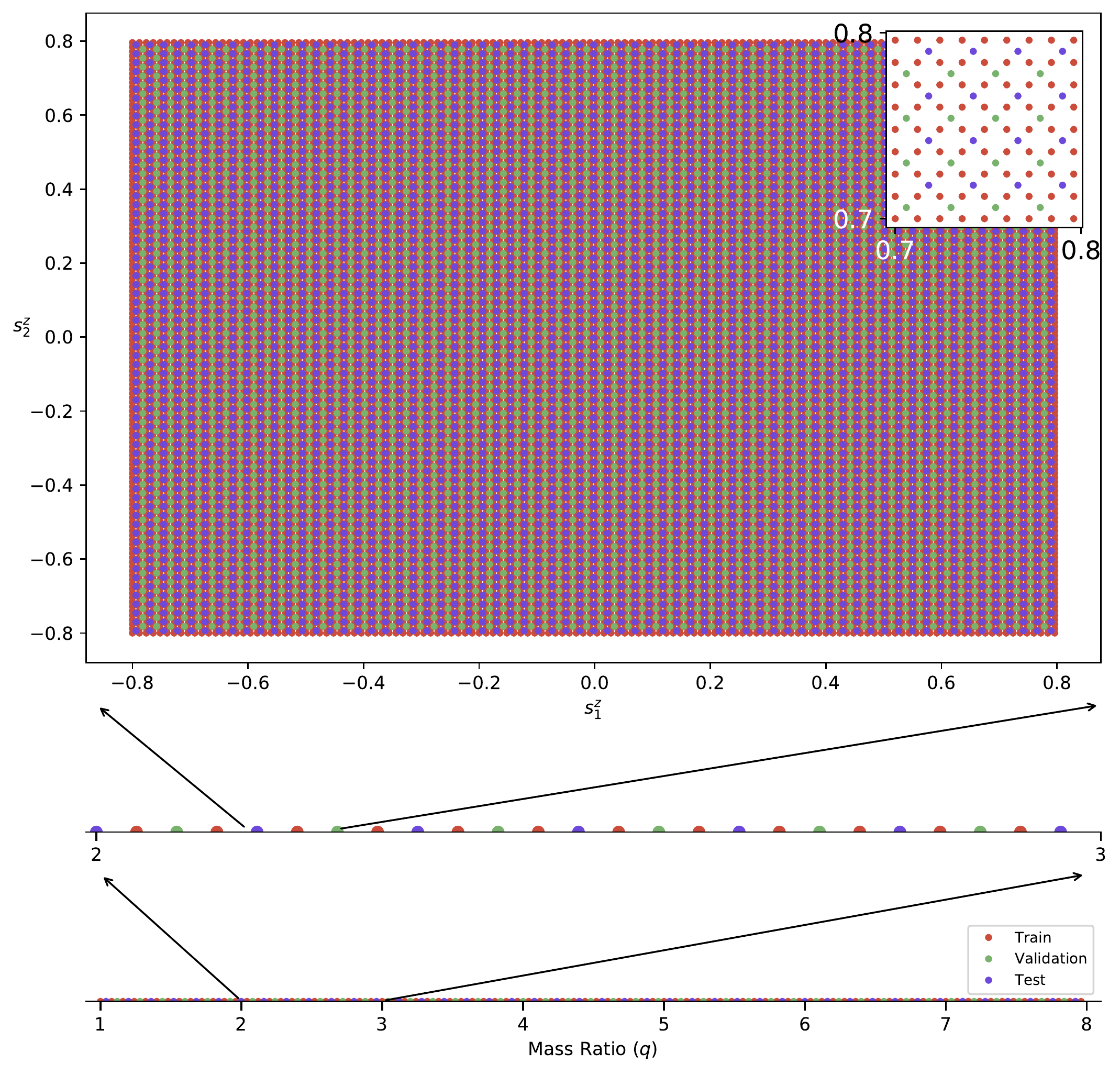}
} 
\caption{Sampling of the signal manifold $q\in[1,\,8]$, $s^z_{\{1,\,2\}}\in[-0.8,\,0.8]$ to construct the training (light blue dots), validation (dark blue dots) and testing (red dots) data sets. \texttt{This Figure was produced by the authors of this chapter in the published article~\cite{KHAN2020135628}.}}
\label{fig:data_distribution}
\end{figure}

The model introduced in~\cite{KHAN2020135628} was trained, 
validated and tested with \(\ell=|m|=2\) waveforms produced 
with the \texttt{NRHybSur3dq8}~\cite{PhysRevD.99.064045} surrogate
model. These signals cover a time span \(t\in[-10,000\,\textrm{M}, 
130\,\textrm{M}]\) with a time step \(\Delta t = 0.1\,\textrm{M}\). 
The training set is generated by sampling the mass ratio 
$q\in[1,8]$ in steps of \(\Delta q = 0.08\); and the individual 
spins  $s^z_i\in[-0.8, 0.8]$ in steps of \(\Delta s^z_i = 0.012\). 
This is equivalent to $\sim1.5$ million waveforms. The validation 
and test sets are generated by alternately sampling the 
intermediate values, i.e. by sampling $q$ and $s^z_i$ in steps of 
$0.16$ and $0.024$ to lie between the training set values, for a total 
of $\sim 190,000$ waveforms each, respectively. The distributions 
of parameters for training, validation and test sets is shown 
in Fig~\ref{fig:data_distribution}. The entire data set is 
$\sim1.5$TB in size, and  
\texttt{mpi4py} is used to parallelize data generation.

\begin{figure}
    \centering
    \includegraphics[width=\linewidth]{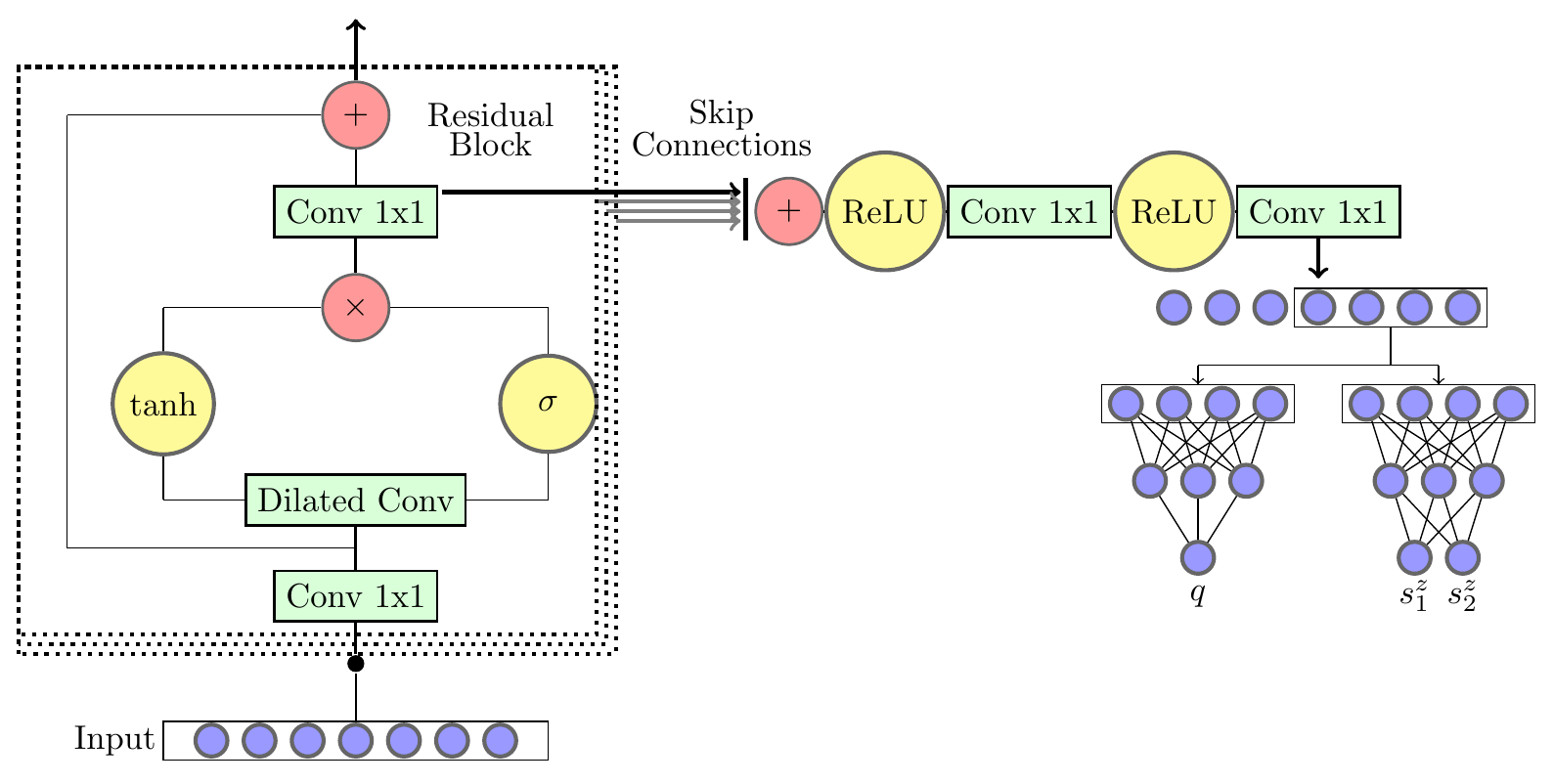}
    \caption{Physics-inspired neural network architecture. Residual blocks of \texttt{Wavenet} (left panel) and leaf layers (right panel).  \texttt{This Figure was produced by the authors of this chapter in the published article~\cite{KHAN2020135628}.}}
    \label{fig:architecture}
\end{figure}

\noindent\textbf{Neural network architecture} The neural network 
architecture consists of two fundamental components, a shared root 
consisting of layers slightly modified from the 
\texttt{WaveNet}~\cite{2016wavenet} architecture, and two branches 
consisting of fully connected layers that take in features extracted 
from the root to predict the mass ratio and the individual spins of 
the  binary components, respectively, as illustrated in Fig~\ref{fig:architecture}. 

\noindent\textbf{Physics-inspired optimization scheme} The 
effective one-body Hamiltonian for moderately spinning black 
holes, 

\begin{equation}
    S_{\textrm{eff}} = \sigma_1 s^z_1 + \sigma_2 s^z_2\,,
    \label{eq:eff_spin}
\end{equation}

\noindent where \(\sigma_1 \equiv 1 + \frac{3}{4q}\) and \(\sigma_2 \equiv 1 + \frac{3q}{4}\), and the effective spin parameter

\begin{equation}
    \sigma_{\textrm{eff}} = \frac{m_1s^z_1 + m_2 s^z_2}{m_1 + m_2} = \frac{q s^z_1 + s^z_2}{1 + q}\,,
    \label{eq:alt_eff_spin}
\end{equation}

\noindent were used to train the model and provide tight constraints 
for the individual spins \(s^z_i\). The performance of the 
neural network was assessed by computing the overlap 
${\cal{O}}\left(h,\,s\right)$, between every waveform in the testing 
dataset, \(h(\theta^i)\) with ground-truth parameters 
\(\theta^i\), and the signal, \(s\), that best 
describes \(h\) according to the neural network model, i.e., 
\(s(\hat{\theta}^i)\) using the relation

\begin{equation}
\label{over}
{\cal{O}}  (h,\,s)= \underset{ t_c\, \phi_c}{\mathrm{max}}\left(\hat{h}|\hat{s}_{[t_c,\,  \phi_c]}\right)\,,\quad{\rm with}\quad \hat{h}=h\,\left(h | h\right)^{-1/2}\,.
\end{equation}

\begin{figure*}
\centerline{
\includegraphics[width=\linewidth]{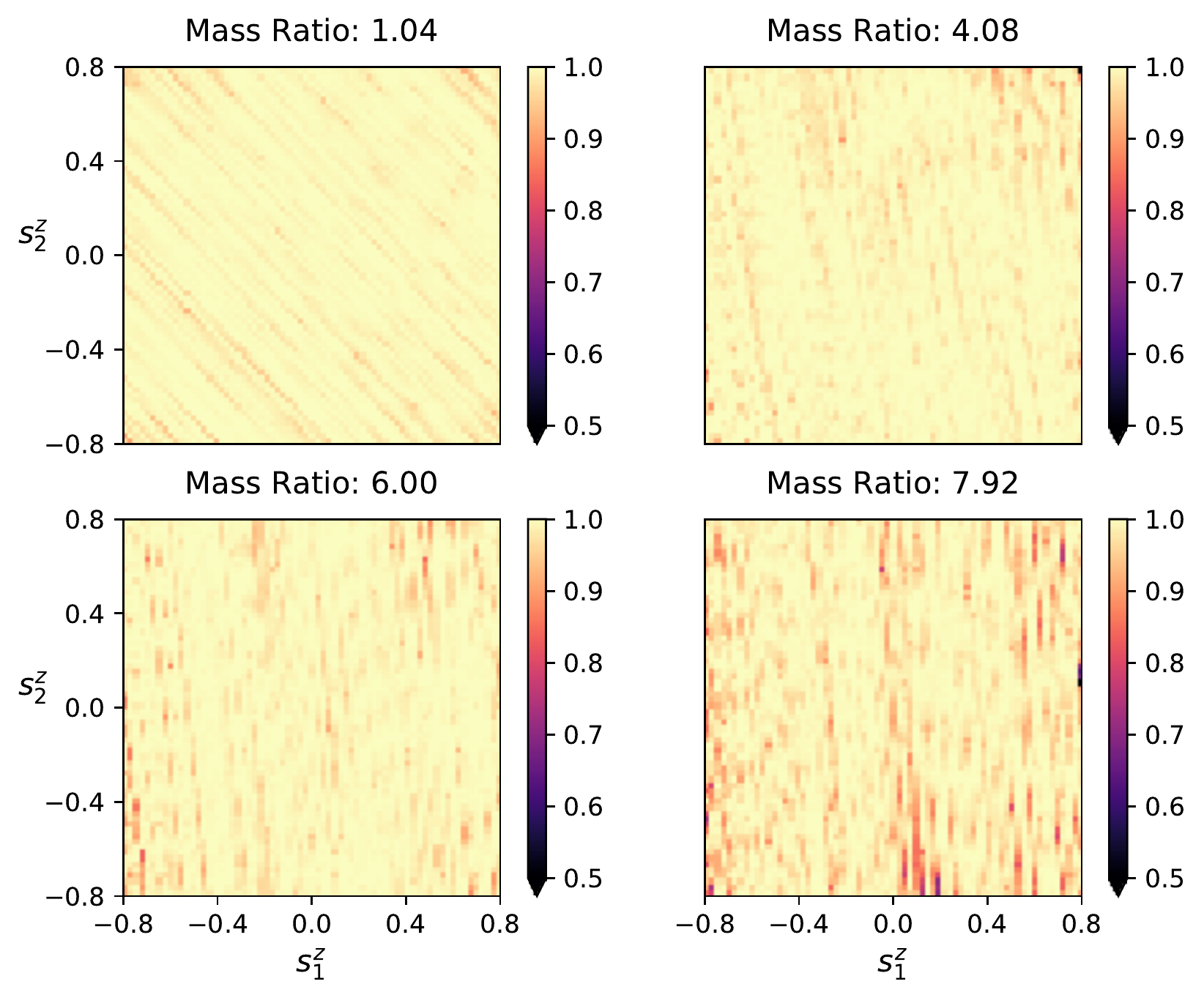}
} 
\caption{Each point in the top and middle panels represents the overlap between a signal in the testing data set and its counterpart whose individual spins and mass ratio are predicted by the neural network model. The mass ratio slices presented in this figure were randomly selected from the testing data set.  \texttt{This Figure was produced by the authors of this chapter in the published article~\cite{KHAN2020135628}.}}
\label{fig:wf_match_imshow}
\end{figure*}

\noindent Figure~\ref{fig:wf_match_imshow} indicates that deep 
learning accurately measures the mass ratio and individual 
spins over a broad range of the parameter space under 
consideration.

This study~\cite{KHAN2020135628} has 
shown that while vanilla neural networks provide uninformative 
predictions for the astrophysical parameters of black hole 
mergers, physics-inspired models provide accurate 
predictions. Thus, these approaches may be investigated in 
the context of parameter estimation to further constrain existing 
measurements for the spin distribution of observed events.

As in the case of gravitational wave detection, deep learning applications for inference are progressing at a very rapid pace. The extension of existing neural network models to characterize real signals with a broader range of reported signal-to-noise ratios, in particular at the low end, will mark 
a major milestone on this exciting front. 

The importance of developing novel signal processing tools and computing approaches is underscored by the computing needs of established, though poorly scalable and compute-intensive algorithms, which burned about 500M CPU core-hours in astrophysical searches, follow up studies and detector characterization analyses during the third observing run. Furthermore, the second observing run indicates that about 10M CPU core-hours of computing were needed for \({\cal{O}}\left(10\right)\) detected events. In the scenario of a third generation gravitational wave detection network with three interferometers, the number of observed events per year may be of order \({\cal{O}}\left(10^3\right)\), and thus the computing needs will grow by 3 orders of magnitude. In brief, it is essential to pursue innovation in signal processing tools, computing methodologies and hardware architectures if we are to realize the science goals of gravitational wave astrophysics~\cite{scenarioligo:2016LRR}.

\section{\textit{Deep Learning for the classification and clustering of noise anomalies in gravitational wave data}}
\label{sec:classification}

While deep learning is now customarily used to extract information 
from complex, noisy, and heterogeneous datasets, it is worth exploring 
and removing known sources of noise from experimental datasets. 
This is particularly relevant in the context of gravitational 
wave astrophysics, since noise anomalies---or glitches---tend to 
contaminate and even mimic real gravitational wave signals. 

\begin{figure*}
	\centering     
	{\includegraphics[width=.9\textwidth]{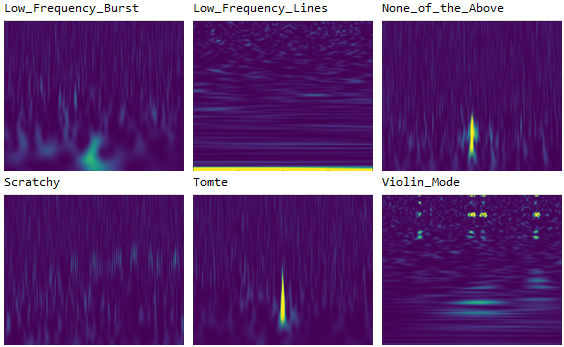}}		
	\caption{Sample of glitches in the \texttt{Gravity Spy} dataset from advanced LIGO's first observing run. \texttt{This Figure was produced by the authors of this chapter in the published article~\cite{glitch_clustering}.}}
	\label{fig:glitches}
	\vspace*{-6mm}
\end{figure*}

The \texttt{Gravity Spy} project aims to identify, classify and 
excise instrumental and environmental sources of noise that decrease the sensitivity of ground-based 
gravitational wave detectors~\cite{grav_spy}. A sample of 
glitches classified by the \texttt{Gravity Spy} are shown in Figure~\ref{fig:glitches}.

\noindent As citizen science efforts continue to increase the 
number of glitches classified through \texttt{Gravity Spy}, 
it may be possible to automate their classification, or to utilize human-in-the-loop machine learning methods. An initial approach 
for glitch classification was presented in~\cite{grav_spy}. 
This method 
consisted of using the small and unbalanced dataset of \texttt{Gravity Spy} glitches to train a neural network model from the 
ground up. A method for automatic glitch 
classification was introduced in~\cite{glitch_clustering}. This approach combined 
deep learning with transfer learning for glitch classification. 
Specifically, this study showed that models such as \texttt{Inception}, \texttt{ResNet-50}, and \texttt{VGG} that have been pre-trained 
for real-object recognition using \texttt{ImageNet} may be fine-tuned through transfer learning to enable optimal classification of small and unbalanced 
datasets of spectrograms of glitches curated by \texttt{Gravity Spy}. 
This approach provided state-of-the-art classification accuracy, 
reduced the length of the training stage by several orders of magnitude, and eliminated the need for hyperparameter optimization. More importantly, 
both \texttt{ResNet-50} and \texttt{Inceptionv3} achieved a classification accuracy of 98.84\% on the test set despite being trained independently via different methods on different splits of the data, and obtained 100.00\% accuracy when considering the top five predictions. This means that for any given input, the true class can be narrowed down to within five classes with 100.00\% confidence. This is particularly useful, since the true class of a glitch is often ambiguous, even to human experts. Finally, this study~\cite{glitch_clustering} also showed that neural 
networks may be truncated and used as feature extractors for unsupervised clustering to automatically group together new classes of glitches and noise anomalies. 

Other techniques, such as multi-view convolutional neural networks~\cite{2017arXiv170500034B} and similarity learning~\cite{2019PhRvD..99h2002C}, 
have also been explored to automate glitch classification~\cite{2017arXiv170500034B}. Discriminative embedding functions have also been explored to cluster glitches according to their morphology~\cite{2018arXiv180502296B}. The use of machine learning 
to identify glitches by gathering information from environmental and 
detector data channels has also been reported in~\cite{2020PhRvD.101j2003C}.

The development of a framework to enable online identification 
of simulated glitches was introduced in~\cite{2018CQGra..35i5016R}. 
The wealth of curated data from \texttt{Gravity Spy} may soon enable 
online data quality studies, providing timely and critical input for 
low-latency gravitational wave detection and parameter estimation  
analyses.

\section{\textit{Deep Learning for the construction of Galaxy Catalogs in Large Scale Astronomy Surveys to enable gravitational wave standard-siren measurements of the Hubble constant}}
\label{sec:galaxy}

The previous sections have summarized the state-of-the-art 
in deep learning applications for gravitational wave astrophysics. 
Deep learning is also being investigated in earnest to address 
computational grand challenges for large-scale electromagnetic 
and neutrino surveys~\cite{Nat_Rev_2019_Huerta}. 

This section showcases how to combine deep learning, 
distributed training and scientific visualization to automate 
the classification of galaxy images collected by different surveys. 
This work is timely and relevant in preparation for the next 
generation of electromagnetic surveys, which will significantly 
increase survey area, field of view, and alert production, 
leading to unprecedented volumes of image data and catalog sizes. 
Furthermore, since gravitational wave observations enable a 
direct measurement of the luminosity distance to their 
source~\cite{Schutz:1986Nature}, they may be used in 
conjunction with a catalog of potential host galaxies to 
establish a redshift-distance relationship and measure the 
Hubble constant. This has already been demonstrated in practice 
with the neutron star merger GW170817, whose 
electromagnetic counterpart allowed an unambiguous 
identification of its host galaxy~\cite{GWH:NaturA}. Compact binary mergers 
without electromagnetic counterparts, such as the black hole merger GW170814, have been combined with galaxy catalogs provided by the Dark Energy Survey Year 3~\cite{Soares-Santos:2019irc} to estimate the Hubble 
constant. Therefore, to enable these type of statistical analysis 
it is necessary to automate the construction of complete 
galaxy catalogs~\cite{asad:2018K} .

A method to automate galaxy classification was introduced in~\cite{asad:2018K}. The basic idea consists of leveraging the 
human-labelled galaxy images from the Galaxy Zoo project to fine-tune 
a neural network model that was originally pre-trained for real-object 
recognition using \texttt{ImageNet}. As described in~\cite{asad:2018K}, 
this approach not only enabled state-of-the-art  
classification for galaxies observed by the Sloan Digitial Sky 
Survey (SDSS), but also for Dark Energy Survey (DES) galaxies. A fully trained model 
can classify about 10,000 test images within 10 minutes 
using a single Tesla P100 GPU.

\begin{figure*}[h!]
\centerline{
\includegraphics[width=0.45\linewidth]{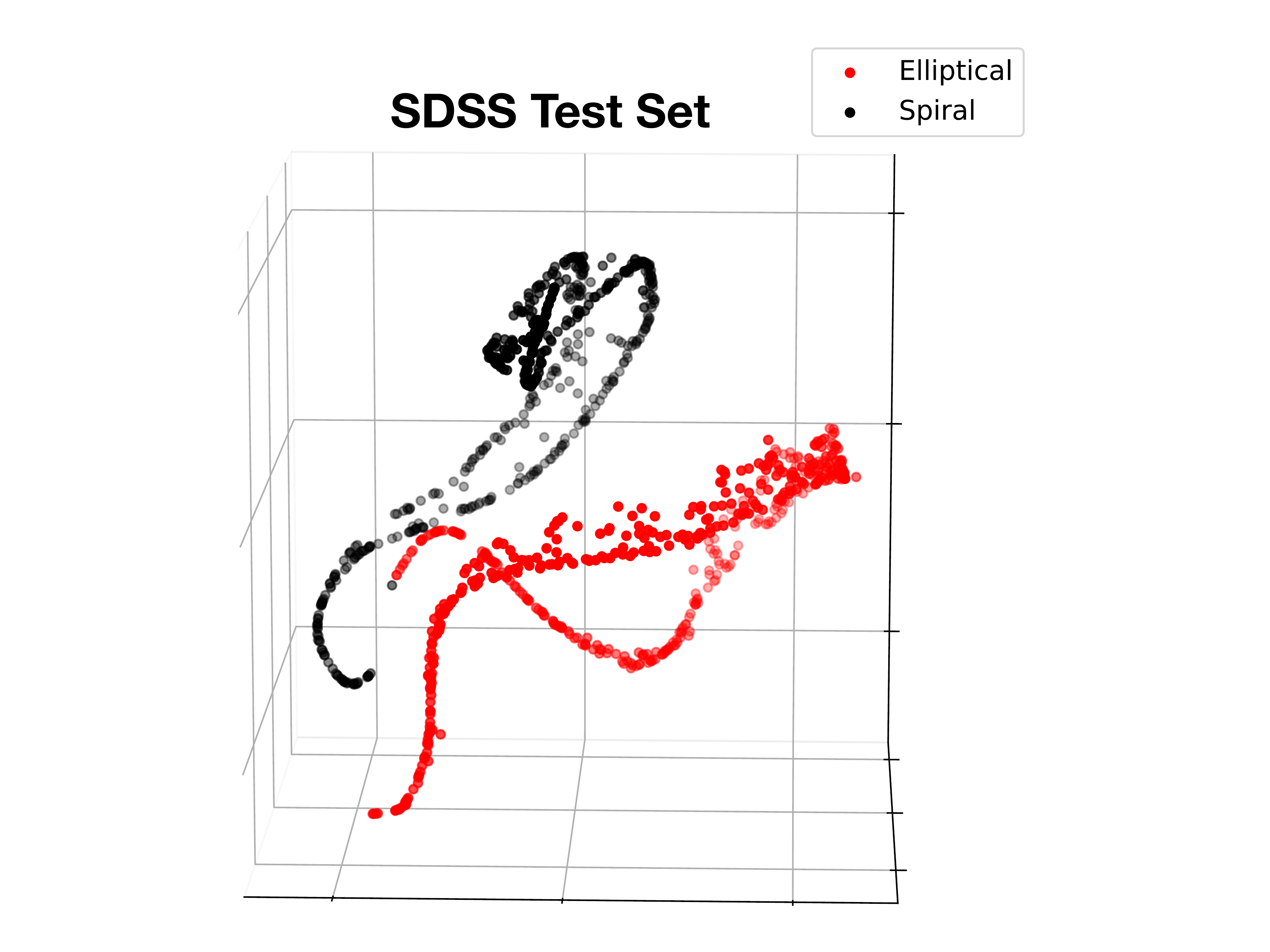}\hspace{-3.5em}%
\includegraphics[width=0.45\linewidth]{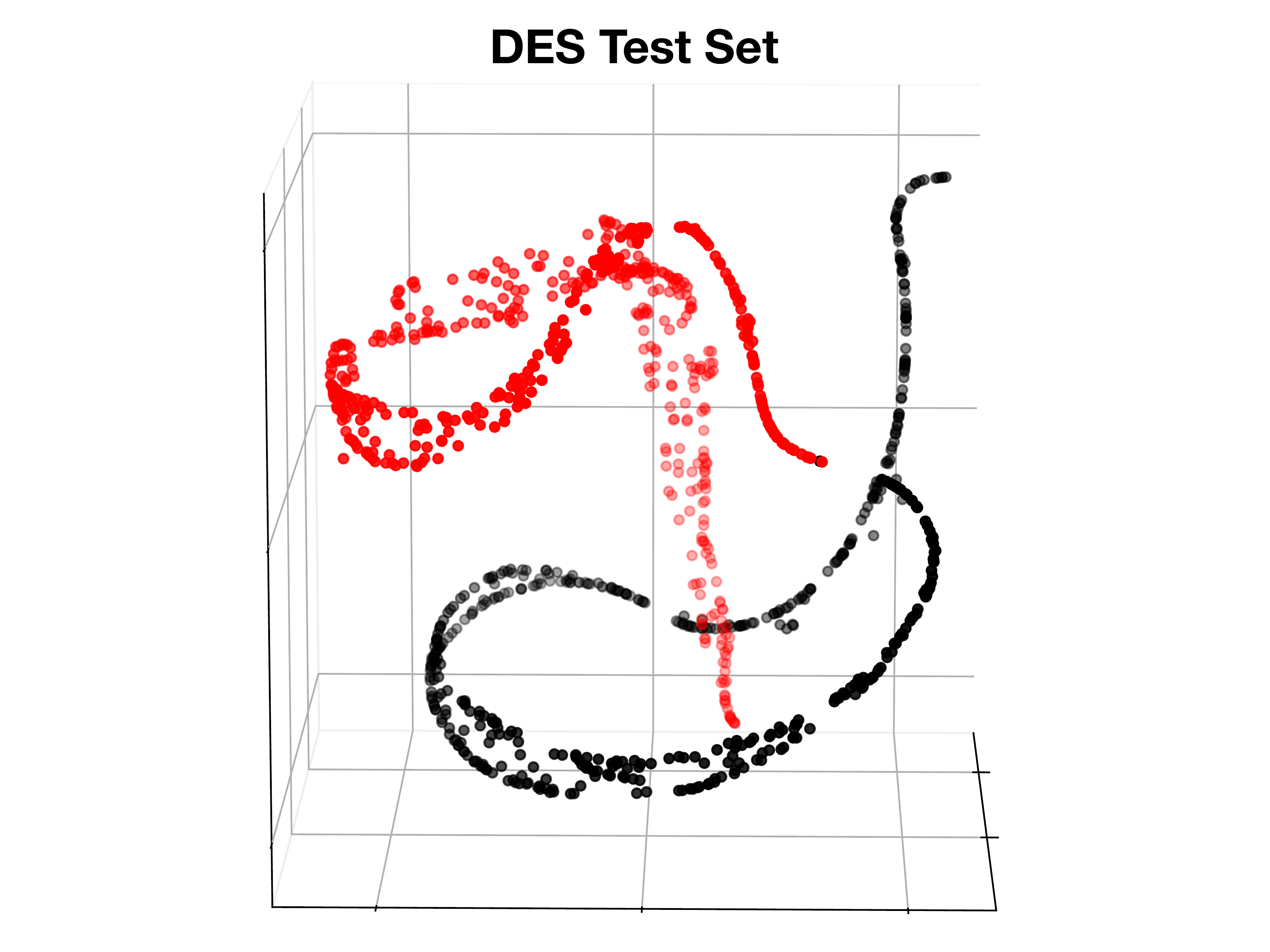}\hspace{-3.5em}%
\includegraphics[width=0.45\linewidth]{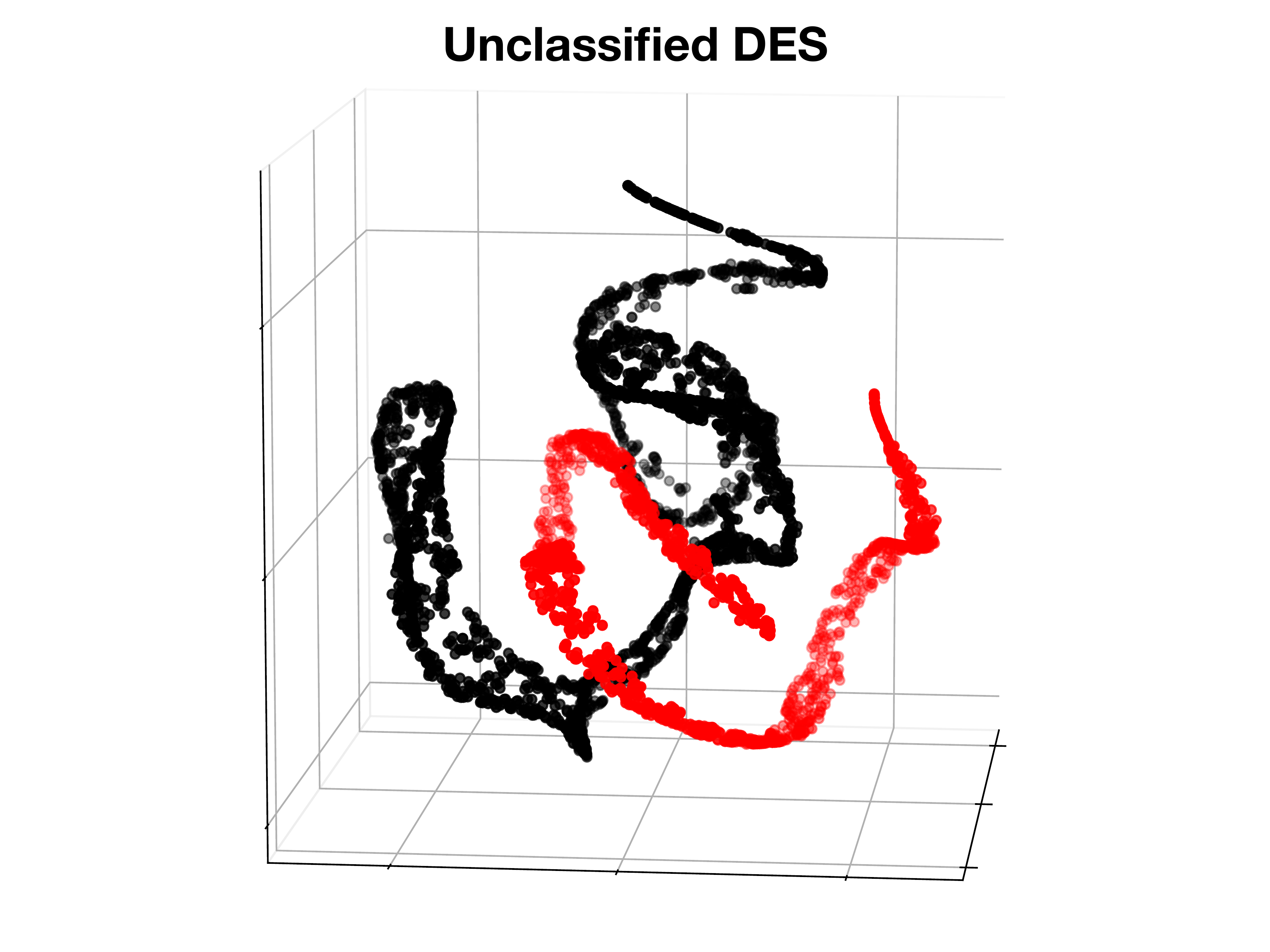}
}
\caption{t-SNE visualization of the clustering of SDSS and DES test sets, and unlabelled DES test.  \texttt{This Figure was produced by the authors of this chapter in the published article~\cite{asad:2018K}.}}
\label{fig:tSNE}
\end{figure*}

\noindent \textbf{Interpretability} While the complexity of deep 
learning models is a major asset in processing large 
datasets and enabling data-driven discovery, it also poses 
major challenges to interpreting how these models acquire 
knowledge and use said knowledge to make predictions. This 
challenge has been widely recognized, 
and novel exploration techniques as well as visualization 
approaches are required to aid in deep learning model 
interpretability. Figure~\ref{fig:tSNE} present the 
activation maps of the second-to-last layer of the model 
described 
above for automated galaxy classification using t-SNE~\cite{TSNE}. 
t-SNE is a nonlinear dimensionality reduction technique that 
is particularly apt for visualizing high-dimensional datasets 
by finding a faithful representation in a low-dimensional 
embedding, typically 2-D or 3-D. These 3-D projections 
are then visualized at different training iterations, 
and at the end of the training they neatly cluster into 
two groups, corresponding to spirals and elliptical. A 
scientific visualization of this clustering algorithm 
for the entire Dark Energy Survey test set is presented at ~\cite{viztsne2}.

In summary, methods that have been explored elsewhere for 
automated image classification, as in the case of 
glitch classification~\cite{glitch_clustering}, 
may be seamlessly applied to 
galaxy classification~\cite{asad:2018K}. This is an active area in deep learning research, i.e., the development of commodity tools that 
may be used across disciplines.

\section{\textit{Challenges and Open Problems}}

We have seen quite a few successful examples on how AI 
is able to improve the detection, parameter 
estimation, waveform production, and denoising of
gravitational waves in the context of real advanced LIGO 
data. However, there are still important challenges to 
be addressed to turn AI into the preferred signal-processing tool for discovery at scale. 

One major challenge is the huge computational cost for constructing and updating AI models, including exploration of the model architectures, hyper-parameter tuning, and training of 
AI models with streaming simulation/experimental data.  
To improve the convergence of training, we need to develop new initialization and optimization techniques for the network weights. Distributed training is also crucial for processing large simulation and observational data. In the section below, we detail our vision for combining AI and high performance computing for Multi-Messenger Astrophysics. We anticipate that in the future, we will have pre-trained AI models available in large-scale scientific projects, e.g., LIGO, LSST, SKA, etc., for 
production scale classification and inference. It is important to incorporate computing at different levels, e.g., edge computing for real-time on-site prediction and cloud computing for updating the models with new training data to adapt to the changes in the sensitivity of detectors, and physical parameter space.   

The AI models discussed in this chapter are trained with particular noise statistics. Previous works~\cite{Shen:2019vep, Wei:2019zlc,Wei:2020sfz} have shown that the trained models are robust to small changes in noise statistics. As LIGO and other astronomical observatories continue to enhance their detection capabilities, researchers will uncover new types of noise anomalies that may contaminate or mimic transient astrophysical events. It is then important to develop new unsupervised and semi-supervised learning techniques to tell apart unexpected noise anomalies from real events. Since the noise statistics of 
observatories vary with time, it will be necessary to retrain the network models every few hours, a light-weight computational 
task that may readily completed within a few minutes with cloud 
computing resources. This approach may also be useful to drive 
the convergence of centralized HPC platforms that are 
essential to accelerate the training of AI models from the ground up. Once fully trained, these AI models may be deployed at the edge to enable 
real-time inference of massive, multi-modal and complex datasets generated by scientific facilities. Thereafter, these models may be fine-tuned with 
light-weight, burst-like re-training sessions with cloud computing.

\section{\textit{Convergence of Deep Learning with High Performance Computing: An emergent framework for real-time Multi-Messenger Astrophysics discovery at scale}}

This section provides a vision for the future of deep learning 
in Multi-Messenger Astrophysics. First, deep learning has 
rapidly evolved from a series of disparate efforts into a 
worldwide endeavor~\cite{Nat_Rev_2019_Huerta}. As described 
in the previous sections, there has been impressive progress 
across all 
fronts in gravitational wave astrophysics including detection, 
parameter estimation, data cleaning and denoising, and glitch 
classification. While the vast majority of these approaches have 
used vanilla neural network models, there is an 
emergent trend in which deep learning is combined with domain 
expertise to create physics-inspired architectures and 
optimization schemes to further improve neural network predictions~\cite{KHAN2020135628}. 

Another interesting trend in recent studies is the use of high-dimensional signal manifolds---one of the key considerations that 
led to the exploration of deep learning. Applying deep learning to 
create production-scale data analysis algorithms involves the 
combination of large datasets and distributed training on high 
performance computing platforms to reduce time-to-insight. This 
approach is in earnest development across disciplines, from 
plasma physics to genomics~\cite{2020arXiv200308394H,hep_kyle,ward_blaiszik_foster_assary_narayanan_curtiss_2019,Blatti642124}. 

\begin{figure*}[h!]
\centerline{
\includegraphics[width=\linewidth]{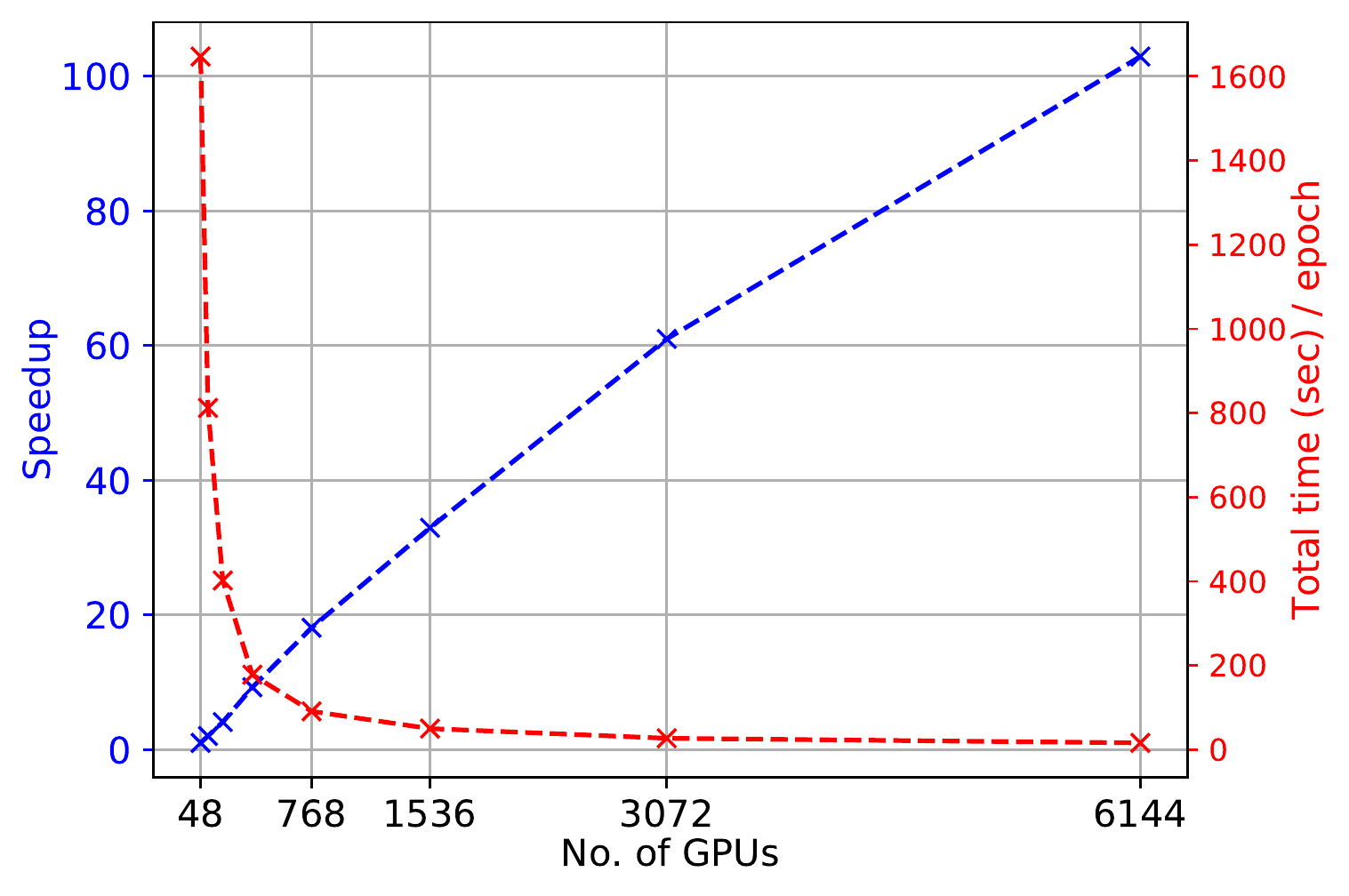}
}
\caption{600-fold speedup in training for a physics-inspired model to characterize the signal manifold of spinning binary black hole mergers. This acceleration is produced by deploying and tuning distributed training schemes on the Summit supercomputer at Oak Ridge National Laboratory. \texttt{This Figure was produced by the authors of this chapter in the published article~\cite{KHAN2020135628}.}}
\label{fig:scaling}
\end{figure*}

Figure~\ref{fig:scaling} shows recent progress using the 
Summit supercomputer to accelerate by 600-fold the training of 
physics-inspired deep learning models for gravitational wave 
astrophysics~\cite{KHAN2020135628}. Mirroring the successful approach 
of corporations that lead innovation in artificial intelligence 
research, projects such as the Data and Learning Hub for 
Science~\cite{dlhub,blaiszik_foster_2019} have provided an 
open source platform to share artificial intelligence models 
and data with the broader 
community. 
This approach will accelerate the development of novel artificial 
intelligence tools to enable breakthroughs in science and 
technology in the big data era.

In addition to combining artificial intelligence and extreme-scale 
computing to reduce time-to-insight, there is an ongoing effort to 
incorporate artificial intelligence into the software stacks used to 
numerically simulate multi-scale and multi-physics processes, such 
as neutron star mergers~\cite{2020PhRvD.101h4024R}. Through these approaches, it may be 
feasible to accurately capture the physics of subgrid scale processes 
such as turbulence at a fraction of the time and computational 
resources currently needed for high-quality simulations. 
In essence, this is 
promoting artificial intelligence as a guiding tool to 
maximize the use and reach of advanced cyberinfrastructure facilities.

Finally, as artificial intelligence and innovative computing 
become widely adopted as the go-to signal processing 
and computing paradigms, it is essential to not become complacent 
in the quest for better signal processing tools. Since the development 
of artificial 
intelligence goes well beyond academic pursuits, it will be 
important to keep transferring and cross-pollinating innovation 
between academia, industry and technology. At the same time, 
it is essential to keep pursuing translational 
research, e.g., how to reuse algorithms for real-object 
recognition in the context of glitch classification~\cite{glitch_clustering} 
and galaxy classification~\cite{asad:2018K}, or how to adapt and combine algorithms for gravitational wave 
denoising~\cite{Wei:2019zlc} and earthquake 
detection~\cite{Perole1700578} to other tasks, like the identification of heart conditions~\cite{arjun_cardio}. 

The future of artificial 
intelligence and innovative computing for Multi-Messenger 
Astrophysics 
is in the hands of bold innovators that will continue to expand the frontiers of discovery.

\section{\textit{Cross-References}}

This chapter is related to the following entries in this book:

\vspace{3mm}
\begin{itemize}
\item \noindent \textit{Introduction to gravitational wave astronomy} by Nigel Bishop\\
\item\noindent\textit{Binary neutrons stars} by Luca Baiotti\\
\item\noindent\textit{Dynamic formation of stellar-mass binary black holes} by Bence Kocsis\\
\item\noindent\textit{Multi-messenger astronomy} by Marica Branchesi \\
\item\noindent \textit{Numerical Relativity for gravitational wave source modeling} by Zhoujian Cao\\
\item\noindent\textit{Gravitational wave signal detection techniques} by Kipp Cannon\\
\item\noindent \textit{Gravitational wave data characterization and machine learning techniques} by Elena Cuoco
\end{itemize}

\section{\textit{Acknowledgements}}

EAH and ZZ gratefully acknowledge National Science Foundation awards 
OAC-1931561 and OAC-1934757, Department of Energy award DE-SC0021258. 
EAH also acknowledges the Innovative and Novel Computational Impact on Theory and Experiment (INCITE) 
award ``Multi-Messenger Astrophysics at Extreme Scale in Summit''. 
This research used resources of the Oak Ridge Leadership Computing Facility, which is a DOE Office of Science User Facility supported under Contract DE-AC05-00OR22725. 
We thank Roland Haas, Sarah Habib, 
Maeve Heflin, Xiaobo Huang, Asad Khan, Shawn Rosofsky, Hongyu 
Shen, Minyang Tian and William Wei for their contributions writing and editing this chapter.

\clearpage
\bibliography{book_references}
\end{document}